\PassOptionsToPackage{table,prologue}{xcolor}
\documentclass[sigconf]{acmart}

\AtBeginDocument{%
  }

\copyrightyear{2025}
\acmYear{2025}
\setcopyright{cc}
\setcctype{by-nc}
\acmConference[MICRO '25]{58th IEEE/ACM International Symposium on Microarchitecture}{October 18--22, 2025}{Seoul, Republic of Korea}
\acmBooktitle{58th IEEE/ACM International Symposium on Microarchitecture (MICRO '25), October 18--22, 2025, Seoul, Republic of Korea}\acmDOI{10.1145/3725843.3756120}
\acmISBN{979-8-4007-1573-0/2025/10}

\usepackage{listings}
\usepackage{url}
\usepackage{multirow}
\usepackage{subfig}
\usepackage{algorithm2e}
\definecolor{bg}{rgb}{0.95,0.95,0.95}
\newcommand{\omod}[1]{\ \mathrm{mod}\ #1}
\newcommand{\bparagraph}[1]{\textbf{#1}}
\usepackage{makecell}

\definecolor{commentteal}{HTML}{4C787A}
\definecolor{typecolor}{HTML}{B53257}
\definecolor{numgray}{HTML}{666666}
\definecolor{keywordblue}{HTML}{1802F4}

\lstdefinestyle{mintedC}{
    language=C,
    basicstyle=\footnotesize\ttfamily,
    frame=none,
    tabsize=4,
    keepspaces=true,
    numbers=left,
    numbersep=-6pt,
    numberstyle=\tiny\color{numgray},
    breaklines=true,
    showstringspaces=false,
    commentstyle=\color{commentteal}\itshape,
    keywordstyle=[1]\color{typecolor}\bfseries,
    morekeywords=[1]{uint64_t,_Bool,int,char,float,double,long,short,void,unsigned,signed,bool,size_t},
    emph={addmod128},
    emphstyle=\color{keywordblue},
    escapeinside={(*@}{@*)},
}

\lstdefinestyle{shellblock}{
  basicstyle=\footnotesize\ttfamily,
  backgroundcolor=\color{bg},
  frame=single, framerule=0pt,
  rulecolor=\color{bg},
  framextopmargin=.3pt,
  framexbottommargin=.3pt,
  framexleftmargin=-3pt,
  framexrightmargin=-3pt,
  numbers=none,
  columns=fullflexible
}

\begin{document}


\title{Towards Closing the Performance Gap for Cryptographic Kernels Between CPUs and Specialized Hardware}


\author{Naifeng Zhang}
\orcid{0009-0004-0190-4041}
\affiliation{%
  \institution{Carnegie Mellon University}
  \city{Pittsburgh}
  \country{USA}
}
\email{naifengz@cmu.edu}

\author{Sophia Fu}
\orcid{0009-0002-2358-3336}
\affiliation{%
  \institution{Carnegie Mellon University}
  \city{Pittsburgh}
  \country{USA}
}
\email{syfu@andrew.cmu.edu}

\author{Franz Franchetti}
\orcid{0000-0002-3529-8973}
\affiliation{%
  \institution{Carnegie Mellon University}
  \city{Pittsburgh}
  \country{USA}
}
\email{franzf@andrew.cmu.edu}




\begin{abstract}
Specialized hardware like application-specific integrated circuits (ASICs) remains the primary accelerator type for cryptographic kernels based on large integer arithmetic. Prior work has shown that commodity and server-class GPUs can achieve near-ASIC performance for these workloads. However, achieving comparable performance on CPUs remains an open challenge. This work investigates the following question: How can we narrow the performance gap between CPUs and specialized hardware for key cryptographic kernels like basic linear algebra subprograms (BLAS) operations and the number theoretic transform (NTT)?

To this end, we develop an optimized scalar implementation of these kernels for x86 CPUs at the per-core level. We utilize SIMD instructions---specifically AVX2 and AVX-512---to further improve performance, achieving an average speedup of 38 times and 62 times over state-of-the-art CPU baselines for NTTs and BLAS operations, respectively. To narrow the gap further, we propose a small AVX-512 extension, dubbed multi-word extension (MQX), which delivers substantial speedup with only three new instructions and minimal proposed hardware modifications. MQX cuts the slowdown relative to ASICs to as low as 35 times on a \textit{single} CPU core. Finally, we perform a roofline analysis to evaluate the peak performance achievable with MQX when scaled across an entire multi-core CPU. Our results show that, with MQX, top-tier server-grade CPUs can approach the performance of state-of-the-art ASICs for cryptographic workloads.
\end{abstract}



\maketitle

\section{Introduction}
\label{sec:intro}

\begin{figure}[t]
\centering
    \includegraphics[width=\columnwidth,trim={15mm 0mm 15mm 0mm},clip]{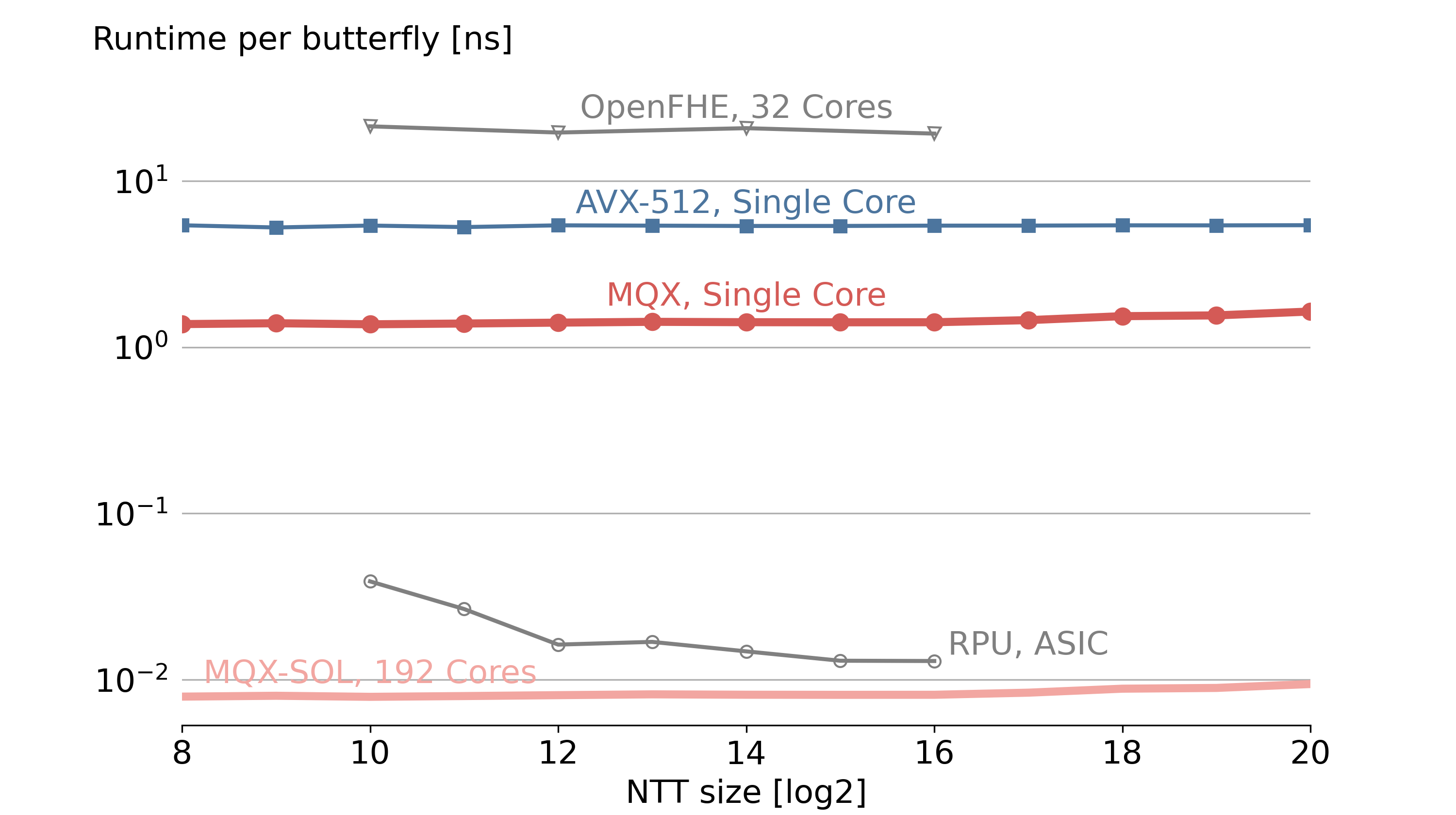}
    \caption{Performance comparison of NTT implementations on CPUs and an ASIC (lower is better). Our implementation of NTT using AVX-512 on a single core of AMD EPYC 9654 achieves a 3.8 times speedup compared to the state-of-the-art FHE library~\cite{al2022openfhe}, which is benchmarked on AMD EPYC 7502 with 32 cores~\cite{soni2023rpu}. When scaled to 192 cores of AMD's server-grade CPU, the peak performance achievable with MQX suggests that CPU-based implementations could approach near-ASIC performance for NTTs.}
    \label{fig:teaser}
    \vspace{2mm}
\end{figure}

As today's computing landscape shifts towards distributed and cloud computing environments, advanced encryption techniques, such as fully homomorphic encryption (FHE), are gaining interest for privacy-preserving computations. FHE keeps the data encrypted as soon as it leaves the local party by allowing computations on the encrypted data on the external side. 
However, despite its strong privacy guarantees, FHE has not seen widespread adoption due to its prohibitive computational overhead, which primarily arises from polynomial arithmetic involving very large coefficients. To handle large integer arithmetic, prior works employ the residue number system (RNS) to decompose very large coefficients (e.g., over 1,000 bits) into smaller components (i.e., residues) that fit within machine words, such as 64 bits. 
Recently, several studies have explored the use of 128-bit residues~\cite{soni2023rpu,zhang2025code,zhou2024fully,nabeel2023cofhee,al2022openfhe}---going beyond common machine word sizes---to reduce the computational overhead associated with modulus raising and reduction~\cite{cheon2017homomorphic}, as well as to reduce the frequency of \textit{bootstrapping}, a highly computationally intensive process in FHE~\cite{al2023demystifying}. 
These works have demonstrated efficient implementations of 128-bit arithmetic on both ASICs~\cite{soni2023rpu,zhou2024fully,nabeel2023cofhee} and GPUs~\cite{zhang2025code}. By natively supporting 128-bit operations in hardware, ASICs can achieve up to a 1,485 times speedup over the OpenFHE~\cite{al2022openfhe} CPU baseline, as shown by RPU~\cite{soni2023rpu}. 
On GPUs, MoMA~\cite{zhang2025code} introduces \textit{multi-word modular arithmetic}, which recursively decomposes large integer arithmetic into machine-word operations. This approach enables commodity GPUs such as NVIDIA GeForce RTX 4090 to achieve near-ASIC performance.
On CPUs, OpenFHE~\cite{al2022openfhe} stands as one of the leading cryptographic libraries for FHE implementations. OpenFHE offers two options for large integer arithmetic: a built-in mathematical backend and the widely-used multi-precision library GMP~\cite{granlund1996gnu}. 
However, both options deliver significantly lower performance compared to ASIC and GPU implementations as discussed above. Therefore, a competitive CPU-based implementation for FHE workloads remains an open challenge.

This paper aims to narrow this performance gap between CPUs and specialized hardware for two core cryptographic kernels in FHE: basic linear algebra subprograms (BLAS) operations and number theoretic transforms (NTTs). NTTs are among the most computationally intensive kernels in FHE, accounting for over 90\% of runtime in many FHE-based workloads~\cite{fan2023tensorfhe}. BLAS operations, likewise, serve as fundamental building blocks for most FHE schemes~\cite{al2022openfhe}. 
We begin by deriving highly optimized scalar implementations (i.e., standard C targeting x86-64 architectures) of NTT and BLAS operations and then explore vector parallelism enabled by single instruction multiple data (SIMD) instructions. As shown in Figure~\ref{fig:teaser}, using Advanced Vector Extension 512 (AVX-512) instructions on a \textit{single core} of AMD EPYC 9654, we can achieve a 3.8 times speedup over OpenFHE that is benchmarked on a 32-core CPU. 

To further enhance performance, we propose a small instruction set architecture (ISA) extension, named multi-word extension (MQX), that augments AVX-512 with just three new vector instructions.
MQX is designed to minimize hardware engineering efforts when integrated with existing AVX intrinsics, drawing on parallels with x86 scalar operations and prior SIMD instructions from Intel’s Knights Corner (KNC) intrinsics~\cite{chrysos2014intel,intel2013guide} and Larrabee New Instructions (LRBni)~\cite{seiler2008larrabee}. 
We define MQX at the ISA level and model its performance using a novel technique, performance projection using proxy ISA (PISA). Through PISA, the measured projected performance of MQX delivers an additional 3.7 times speedup over our AVX-512 baseline for NTTs on AMD EPYC 9654. Finally, we conduct a roofline analysis to estimate the upper performance bound when MQX is optimally scaled across a high-end multi-core AMD CPU. Our analysis shows that the projected performance approaches that of ASIC implementations.

We believe that narrowing the performance gap between CPUs and specialized hardware for FHE has the potential to significantly broaden its applicability and make privacy-preserving applications more accessible. 
With this reduced gap, all FHE computations could remain on the CPU, eliminating the need for GPU- or ASIC-based acceleration. 
This would also remove the substantial overhead incurred by data transfers between separate memory spaces, a critical performance bottleneck for FHE at the application level~\cite{samardzic2021f1}.
Consequently, our approach gains a strong competitive edge, especially since many GPU and ASIC implementations reported in the literature do not account for data transfer times in their performance evaluations of cryptographic kernels. 

\newpage
\bparagraph{Contributions.} This paper makes the following contributions: 
\begin{enumerate}
    \item Efficient implementations of cryptographic kernels on CPUs using scalar, AVX2, and AVX-512 instructions.
    \item A proposed ISA extension---multi-word extension (MQX)---that addresses the performance bottlenecks of AVX-512 for large integer arithmetic and further improves the performance of CPU-based cryptographic kernels.
    \item A demonstration that, on a single CPU core, AVX-512-based NTTs and BLAS operations achieve a 38 times and a 62 times speedup, respectively, over the state-of-the-art CPU baselines. The speedup is further compounded by MQX, resulting in a 77 times and 104 times speedup, respectively. Using a \textit{single} CPU core, MQX narrows the performance gap to ASICs to as low as a 35 times slowdown for NTTs.
    \item A roofline analysis of MQX scaling across a multi-core CPU, which suggests that when all cores of a state-of-the-art server-grade CPU are optimally utilized, cryptographic kernels can achieve performance comparable to specialized hardware.
\end{enumerate}

\begin{table*}[t]
\small
    \caption{Implementations of addition with carry in cryptographic settings: scalar, AVX-512, and MQX.}
    \centering
    \label{tab:adc}
    \begin{tabular}{p{0.25\linewidth} p{0.37\linewidth} p{0.28\linewidth}}
    \toprule[1pt]
    \begin{normalsize}\textbf{Scalar}\end{normalsize} & \begin{normalsize}\textbf{AVX-512}\end{normalsize} & \begin{normalsize}\textbf{MQX}\end{normalsize} \\
    \midrule
\begin{minipage}[t]{\linewidth}
\vspace{-3mm}
\begin{lstlisting}[columns=fullflexible]
    // In: uint64_t a, b; _Bool ci
    // Out: uint64_t c; _Bool co
    uint64_t t0, t1; _Bool q0, q1;
    t0 = a + b;  
    t1 = t0 + ci;
    q0 = (t1 < a);
    q1 = (t1 < b);
    co = q0 || q1;
\end{lstlisting}
\end{minipage}
& 
\begin{minipage}[t]{\linewidth}
\vspace{-3mm}
\begin{lstlisting}[columns=fullflexible]
    // In: __m512i a, b; __mmask8 ci
    // Out: __m512i c; __mmask8 co
    __m512i t0, t1, one; __mmask8 q0, q1;
    t0 = _mm512_add_epi64(a, b);
    one = _mm512_set1_epi64(1);
    t1 = _mm512_mask_add_epi64 (t0, ci, t0, one);
    q0 = _mm512_cmp_epu64_mask(t1, a, _MM_CMPINT_LT);
    q1 = _mm512_cmp_epu64_mask(t1, b, _MM_CMPINT_LT);
    co = q0 | q1;
\end{lstlisting}
\end{minipage}
\vspace{0.2mm}
&
\begin{minipage}[t]{\linewidth}
\vspace{-3mm}
\begin{lstlisting}[columns=fullflexible]
    // In: __m512i a, b; __mmask8 ci
    // Out: __m512i c; __mmask8 co
    c = _mm512_adc_epi64(a, b, ci, &co);
\end{lstlisting}

\end{minipage}
\\
    \bottomrule[1pt]
    \end{tabular}
\small
\end{table*}

\section{Background}

In this section, we introduce modular arithmetic and double-word arithmetic, which, when combined, form the large integer arithmetic used in cryptographic kernels within the scope of this work. We then describe the cryptographic kernels of interest---NTT and BLAS operations---and their roles in polynomial arithmetic. Lastly, we discuss SIMD vectorization, a key technique for achieving high performance on CPUs. We follow some of the mathematical notations adopted in prior work on multi-word modular arithmetic~\cite{zhang2025code}.

\subsection{Modular Arithmetic} 
\label{sec:mod_arith}

Let $\mathbb{Z}_q$ be an integer ring modulo $q$. We can define the modular arithmetic mathematically as
\begin{equation}
\begin{aligned}
        c &= a + b  &&\mod q, \\
        c &= a - b  &&\mod q, \\
        c &= ab     &&\mod q,
\end{aligned}
\end{equation}
where $a,b,c\in\mathbb{Z}_q$. 

Given that on typical hardware the modulo operation is significantly more costly than basic operations such as addition and multiplication, we can take advantage of the fact that $0 \leq a < q$ and $0 \leq b < q$ for $a,b\in\mathbb{Z}_q$, and propose an algorithm to compute modular addition:
\begin{equation}
\label{eq:saddmod}
    c =
    \begin{cases}
        a + b - q,      & \text{if } (a + b) \geq q, \\[.8mm]
        a + b,          & \text{otherwise.}
    \end{cases}
\end{equation}
A similar strategy can be applied to modular subtraction:
\begin{equation}
\label{eq:ssubmod}
    c =
    \begin{cases}
        a - b + q,      & \text{if } a < b, \\[.8mm]
        a - b,          & \text{otherwise.}
    \end{cases}
\end{equation}
However, for multiplication, the conditional check is no longer effective. Instead, we employ Barrett reduction~\cite{barrett1986implementing}, which uses a pre-computed value $\mu$ to replace the modulo operation with cheaper operations such as binary shift and multiplication. Formally:
\begin{equation}
\label{eq:barrett}
    c = ab - \lfloor ab \mu / 2^k \rfloor q,
\end{equation}
where $\mu = \lfloor 2^k/q \rfloor$, and $\lfloor . \rfloor$ denotes the floor operation. $\mu$ only needs to be computed once for the same modulus $q$. $k$ is chosen to satisfy $2^{k/2} > q$, which guarantees sufficient precision when approximating division using $\mu$.
Barrett reduction is widely used within the cryptography community~\cite{soni2023rpu,kim2020accelerating,ozerk2022efficient,livesay2023accelerating,wang2023he} as it works on general moduli rather than a specialized modulus chosen for application-specific optimizations (e.g, Goldilocks prime~\cite{hamburg2015ed448}). A key aspect of Barrett reduction is that if the target data bit-width in modular arithmetic is $l$ (usually a power of two), the modulus $q$ must have fewer than or equal to $l-4$ bits. This ensures that $\mu$ stays within $l$ bits. For example, for 128-bit integer arithmetic, $q$ must be less than or equal to 124 bits.

\subsection{Double-Word Arithmetic}
\label{sec:dword_arith}

In this work, we define a \textit{word} (or \textit{machine word}) as the largest integer data type that can fit into a general-purpose register. For example, a word has 64 bits on x86-64 architectures. In the context of this work, we refer to the bit-width of interest for FHE applications as a 128-bit \textit{double-word}.
A double-word can be mathematically represented using the following notation:
\begin{equation}
\label{eq:def_dw}
    [x_0, x_1]_{2^{\omega_0}} = x_0 2^{\omega_0}  + x_1 = x,
\end{equation}
where $\omega_0$ is the machine word width. In this work, $\omega_0 = 64$, with $x_0$ representing the higher 64 bits and $x_1$ the lower 64 bits of a 128-bit integer. The double-word arithmetic, however, is applicable to other machine word widths as well.

Now, we can formally introduce double-word integer arithmetic. Let $a = [a_0, a_1]_{2^{\omega_0}} = a_0 2^{\omega_0} + a_1$, $b = [b_0, b_1]_{2^{\omega_0}} = b_0 2^{\omega_0} + b_1$. Double-word addition can be written as 
\begin{equation}
\begin{aligned}
    &[\delta, c_2]_{2^{\omega_0}} = a_1 + b_1, \\
    &[c_0, c_1]_{2^{\omega_0}} = a_0 + b_0 + \delta,
\end{aligned}
\end{equation}
where $c = [c_0, c_1, c_2]_{2^{\omega_0}}$ and $\delta\in\{0,1\}$. 
Table~\ref{tab:adc} presents three concrete implementations of $[c_0, c_1]_{2^{\omega_0}} = a_0 + b_0 + \delta$ in C, which performs addition with carry-in ($\delta$) and carry-out ($c_0$): the scalar version (i.e., standard C targeting x86-64 architectures), the SIMD version using AVX-512, and the version utilizing our proposed ISA extension, MQX, where a single instruction computes addition with carry in a SIMD fashion. 
Note that both the scalar and AVX-512 implementations in Table~\ref{tab:adc} operate only on the higher bits of a 128-bit integer ($a_0$ and $b_0$). This approach leverages the fact that, under Barrett reduction, we work with 124-bit integers or smaller (as opposed to full 128-bit integers), as discussed in Section~\ref{sec:mod_arith}.
Double-word subtraction can be written as 
\begin{equation}
\begin{aligned}
    &c_1 = a_1 - b_1, \\
    &\delta = 
    \begin{cases}
        1,          &\text{if } a_1 < b_1, \\[.8mm]
        0,          &\text{otherwise,}    \\
    \end{cases} \\
    &c_0 = a_0 - b_0 - \delta,
\end{aligned}
\end{equation}
where $c = [c_0, c_1]_{2^{\omega_0}}$. 
The schoolbook double-word multiplication can be written as 
\begin{equation}
\label{eq:muls}
    c = (a_0 b_0){2^{2\omega_0}} + (a_0 b_1 + a_1 b_0){2^{\omega_0}} + a_1 b_1,
\end{equation} 
which involves four single-word multiplications. 
We also explore an alternative multiplication algorithm, the Karatsuba algorithm~\cite{karatsuba1962multiplication}, which reduces the number of single-word multiplications to three and can be written as:
\begin{equation}
\label{eq:mulk}
    c = (a_0 b_0){2^{2\omega_0}} + \big((a_0 + a_1)(b_0 + b_1) - a_0 b_0 - a_1 b_1\big){2^{\omega_0}} + a_1 b_1.
\end{equation}

\subsection{Polynomial Operations and NTT}

Polynomial operations with coefficients that reside in $\mathbb{Z}_q$ are the building blocks (which constitute the prohibitive overhead) in many advanced encryption schemes such as FHE. In this section, we introduce how point-wise polynomial operations can be captured as BLAS operations and how NTT accelerates polynomial multiplication. 

\bparagraph{Point-wise polynomial operations.} Point-wise polynomial addition, subtraction, and multiplication are commonly utilized in FHE schemes~\cite{al2022openfhe}. Let $f$ and $g$ denote two polynomials of degree $n$, where $f = \sum_{i=0}^n a_ix^i$, $g = \sum_{j=0}^n b_jx^j$, and $a_i, b_j\in\mathbb{Z}_q$. Point-wise polynomial operation acts on the corresponding coefficients of two polynomials ($a_i$ and $b_j$ when $i = j$). 

We can represent each polynomial using a vector; for example, $f$ can be written in the vector form as $[a_0,a_1,\ldots,a_{n}]$. Then, we can use the BLAS abstraction~\cite{blackford2002updated} to capture the three aforementioned point-wise polynomial operations as vector addition, vector subtraction, and point-wise vector multiplication. Vector addition and subtraction can be interpreted as variants of a BLAS Level 1 operation named axpy, which is defined as $y = a x + y$,
where $x$ and $y$ are vectors and $a$ is a scalar. Point-wise vector multiplication can be interpreted as a special case of a BLAS Level 2 operation named gemv, which computes a general matrix-vector multiplication. 

\bparagraph{Polynomial multiplication using NTT.}
Using the same definition of $f$ and $g$ as above, we can define the schoolbook polynomial multiplication of $f$ and $g$ in $\mathbb{Z}_q$ as 
\begin{equation}
    f(x)g(x) = \sum_{j=0}^{2n} \bigg( \Big(\sum_{i=0}^{j} a_i b_{j-i}\Big) \bmod q \bigg) x^j,
\end{equation}
where $a_i = b_i = 0$ for $i > n$. 
While the time complexity of the schoolbook polynomial multiplication is $O(n^2)$, NTT reduces the complexity to $O(n\log n)$. Similar to how the discrete Fourier transform converts a signal from the time domain to the frequency domain, NTT can convert a polynomial from its coefficient form (e.g., $f(x) = x^3 + 3x^2 + 2x + 1 \omod 5$) to its evaluation form (e.g., $\{f(1), f(2), f(3), f(4)\}$). Formally, we define an $n$-point NTT as 
\begin{equation}
    y_k = \sum^{n-1}_{j=0} x_j \omega^{jk}_{n} \!\!\!\! \mod q,\quad \ 0 \leq k \leq n-1,
\label{eq:ntt}
\end{equation}
where $x_j$ and $y_k$ are input and output sequences, respectively, and $\omega_n$ is the $n$-th primitive root of unity. Even though NTT effectively reduces the time complexity of the schoolbook algorithm, polynomial multiplication using NTT still accounts for the majority of the computational overhead. Prior work has shown that NTT accounts for more than 90\% of FHE-based application execution time in practice~\cite{fan2023tensorfhe}. 

\subsection{SIMD Vectorization} 

SIMD vector instructions are ISA extensions designed for parallel computation on short vectors of integers and floating-point numbers. These instructions operate on vector registers, typically 128 to 512 bits wide, and divide them into $v$-way vectors of smaller data types (usually ranging from 2-way to 16-way, where \textit{way} refers to the number of elements or lanes in each vector). 
Popular families of SIMD instructions include MMX, Streaming SIMD Extensions (SSE), Advanced SIMD (Neon), and Advanced Vector Extensions (AVX).

In our work, we focus on two SIMD instruction sets in the AVX family: AVX2 and AVX-512. We utilize the largest integer data type supported by AVX, which is 64 bits, for 128-bit integer arithmetic. 
Introduced in 2013, AVX2 supports vector registers of 256 bits, which can be divided into 4 ways, with each element holding a 64-bit integer. 
AVX-512, released in 2016, extends AVX2 by supporting vector registers of 512 bits (e.g., 8-way 64-bit vectors) and mask registers. AVX-512 is a family of extensions that includes specialized instruction sets such as AVX-512 vector neural network instructions (AVX-512 VNNI) for neural networks and AVX-512 Galois field new instructions (AVX-512 GFNI) for Galois fields. While the core AVX-512 foundation (AVX-512F) extension is required across all AVX-512 implementations, the remaining extensions can be implemented independently of one another.
Table~\ref{tab:adc} presents example AVX-512 instructions for 64-bit integers, including 8-way vector addition \verb|_mm512_add_epi64| and 8-way vector comparison \verb|_mm512_cmp_epu64_mask|.

\section{SIMD-Vectorized Cryptographic Kernels}

To narrow the performance gap between CPUs and specialized hardware, we first investigate the performance improvements achievable using existing technology on state-of-the-art CPUs for cryptographic kernels. Our initial step is to develop a high-performance scalar implementation without utilizing any SIMD instructions. We then further enhance performance by leveraging SIMD instructions, specifically AVX2 and AVX-512. 

\subsection{Scalar Double-Word Modular Arithmetic}

\begin{lstlisting}[columns=fullflexible, captionpos=b, float, floatplacement=t, caption={Scalar double-word modular addition without using 128-bit data types for computation.}, label={lst:dadd_scalar}]
    uint128_t addmod128(uint128_t a, uint128_t b, uint128_t m) {
        _Bool a31, a34, a35, a38, b1, c1, c2, i27, ...;
        uint64_t d1, d2, d3, t28, t29, t30, al, ah, bl, ...;
        al = LO64(a); ah = HI64(a); bl = LO64(b); bh = HI64(b);
        ml = LO64(m); mh = HI64(m);
        t30 = al + bl; q1 = (t30 < al); q2 = (t30 < bl);
        c1 = q1 || q2; t28 = ah + bh; t29 = t28 + c1;
        q3 = (t29 < ah); q4 = (t29 < bh); c2 = q3 || q4;
        a31 = (mh < t29); a35 = (mh == t29); a38 = (ml <= t30);
        a34 = (a35 && a38); i27 = (a31 || a34); i28 = c2 || i27;
        d1 = t30 - ml; b1 = !a38; d2 = t29 - mh;
        d3 = d2 - b1; ch = i28 ? d3 : t29; cl = i28 ? d1 : t30;
        return INT128(ch, cl); 
    }
\end{lstlisting}

To efficiently implement the 128-bit arithmetic used in NTT and BLAS operations with 64-bit machine words, we combine modular arithmetic (Section~\ref{sec:mod_arith}) and double-word arithmetic (Section~\ref{sec:dword_arith}) to construct double-word modular arithmetic. 
We leverage the fact that $0 \leq a < q$ and $0 \leq b < q$ to design efficient modular addition and subtraction algorithms, replacing division by the modulus with a conditional check. 
For modular multiplication, we implement two algorithms: one based on the schoolbook multiplication and the other on the Karatsuba algorithm. Both employ Barrett reduction for efficient modulo operation, which replaces the division with less expensive multiplications and shifts.
Most importantly, under the assumption that $q$ is always less than or equal to 124 bits in our work (necessitated by Barrett reduction), much of the branching logic and conditional assignments can be eliminated.

We implement two versions of scalar double-word modular arithmetic, one uses the \sloppy{natively supported double-word representation (i.e., \verb|unsigned __int128| in C) to store} the results of 64-bit arithmetic, while the other only uses the single-word data type (\verb|uint64_t|) for computation. The former is used for benchmarking, as it allows the compiler to exploit specialized assembly instructions such as add with carry for efficient carry propagation. The second implementation is essential for SIMD-vectorized implementations, as it allows for a more natural translation to AVX2 and AVX-512 instructions, where the maximum data type supported for each vector element is 64 bits.
In Listing~\ref{lst:dadd_scalar}, we demonstrate how scalar double-word modular addition is performed using only 64-bit integers for computation. In this code, \verb|LO| is a macro that extracts the lower 64-bit part of a 128-bit integer, while \verb|HI| extracts the higher 64-bit part. \verb|INT128| combines two 64-bit integers together into a 128-bit integer, treating the first input as the higher part and the second as the lower part. 

\subsection{SIMD-Vectorized Double-Word Modular Arithmetic}
\label{sec:simd_doma}

\begin{lstlisting}[columns=fullflexible, captionpos=b, float, floatplacement=t, caption={Double-word modular addition using AVX-512.}, label={lst:dadd_avx_512}]
    void addmod128(__m512i* ch, __m512i* cl, __m512i ah, __m512i al,
                   __m512i bh, __m512i bl, __m512i mh, __m512i ml) {
        __m512i t30, t28, t29, d1, d2, d3; 
        __mmask8 q1, q2, c1, q3, q4, c2, a31, a35, ...;
        t30 = _mm512_add_epi64 (al, bl);  
        q1 = _mm512_cmp_epu64_mask(t30, al, _MM_CMPINT_LT);
        q2 = _mm512_cmp_epu64_mask(t30, bl, _MM_CMPINT_LT);
        c1 = q1 | q2;
        t28 = _mm512_add_epi64 (ah, bh);
        t29 = _mm512_mask_add_epi64 (t28, c1, t28, one);
        q3 = _mm512_cmp_epu64_mask(t29, ah, _MM_CMPINT_LT);
        q4 = _mm512_cmp_epu64_mask(t29, bh, _MM_CMPINT_LT);
        c2 = q3 | q4;
        a31 = _mm512_cmp_epu64_mask(mh, t29, _MM_CMPINT_LT);
        a35 = _mm512_cmp_epu64_mask(mh, t29, _MM_CMPINT_EQ);
        a38 = _mm512_cmp_epu64_mask(ml, t30, _MM_CMPINT_LE);
        a34 = a35 & a38; i27 = a31 | a34; i28 = c2 | i27;
        d1 = _mm512_sub_epi64(t30, ml); b1 = ~a38;
        d2 = _mm512_sub_epi64(t29, mh);
        d3 = _mm512_mask_sub_epi64(d2, b1, d2, one);
        *ch = _mm512_mask_blend_epi64(i28, t29, d3);
        *cl = _mm512_mask_blend_epi64(i28, t30, d1);
    }
\end{lstlisting}

\begin{figure*}[!htb]
\centering
    \includegraphics[width=0.96\textwidth,trim={220mm 40mm 170mm 15mm},clip]{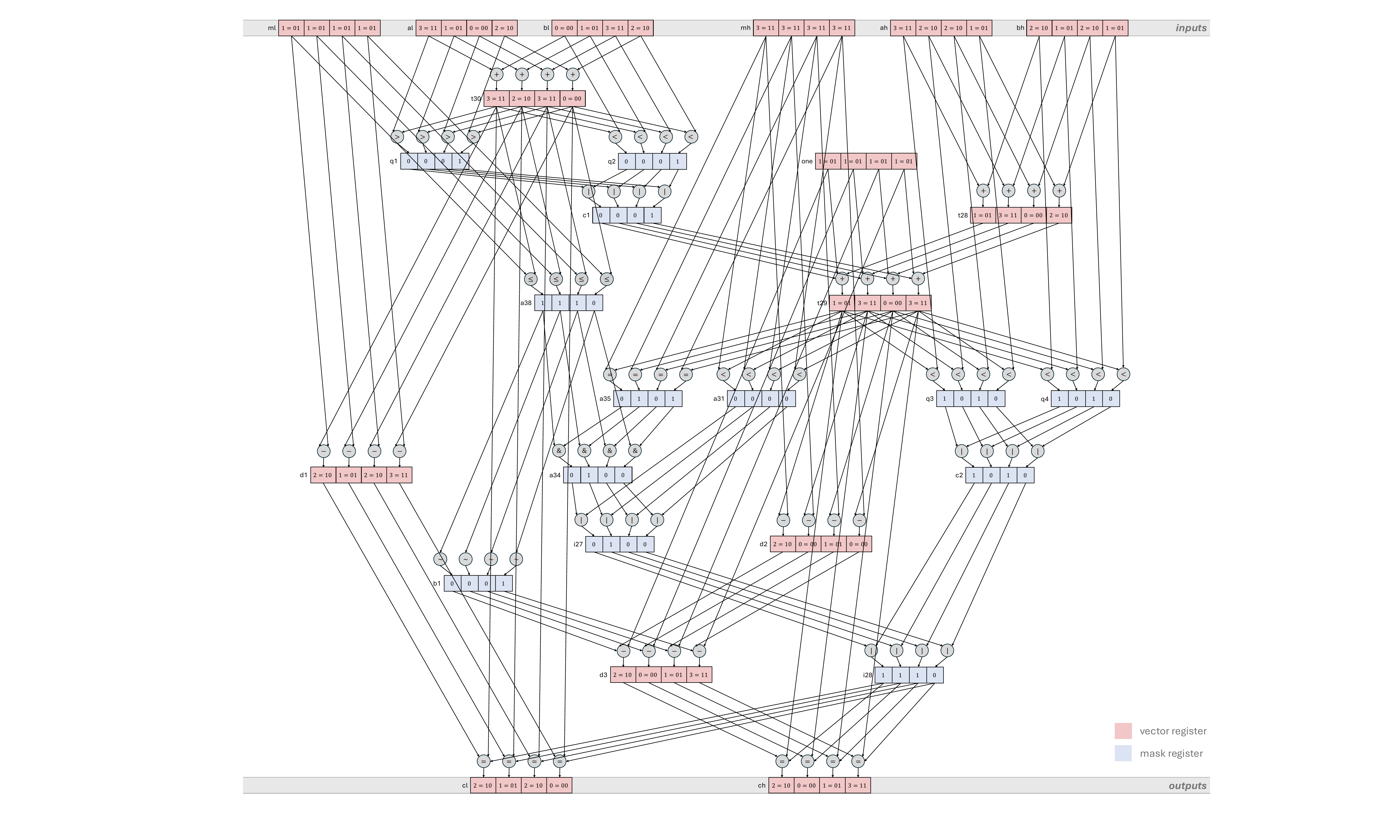}
    \caption{Illustration of double-word modular addition using 4-way vectors, where each element is a 2-bit integer.}
    \label{fig:addmod}
    \vspace{-3mm}
\end{figure*}

To take advantage of the vector parallelism offered by CPUs, we employ SIMD instructions to vectorize the scalar implementation, beginning with AVX-512. Since the largest integer bit-width natively supported by AVX-512 is 64 bits, we divide the 128-bit input vector into two 64-bit vectors: one containing the high parts and the other containing the low parts. This is demonstrated in Figure~\ref{fig:addmod}, where input \verb|a| is split into \verb|ah| and \verb|al|, for example. Consequently, unlike the scalar implementation, which processes one 128-bit integer per input ($a,b,q$ in $c = a + b \omod q$), the vectorized implementation passes in two 512-bit vectors per input, corresponding to eight 128-bit integers, and produces two 512-bit vectors as output.
Figure~\ref{fig:addmod} illustrates our strategy of implementing double-word modular addition using SIMD instructions, with a toy example where each vector contains four elements and each element is 2 bits. This example highlights the complexity of efficiently implementing double-word modular addition using SIMD instructions. 
The corresponding C implementation using AVX-512 is also provided in Listing~\ref{lst:dadd_avx_512}, in which \verb|one| is globally set as \verb|_mm512_set1_epi64(1)|.

When translating from scalar implementation to AVX-512, most arithmetic operations map directly to corresponding AVX-512 instructions that operate on 512-bit vectors. Additionally, AVX-512 supports mask registers and associated operations, which allows us to perform conditional statements from the scalar code without branching. For instance, the vector compare function in AVX-512 outputs a packed 8-bit mask, with each bit corresponding to an element index. This mask can then be used in subsequent operations to control whether an operation is applied to a given index or lane within the vector. This capability allows us to use instructions such as \verb|_mm512_mask_blend_epi64|, which takes two vectors and an 8-bit mask and produces a blended result with values based on which mask bits are set. 

We also implement SIMD-vectorized modular arithmetic using AVX2.
When transitioning from AVX-512 to AVX2, each SIMD instruction handles half as many elements as AVX-512 (four SIMD lanes instead of eight). Moreover, due to the absence of unsigned comparisons and mask registers in AVX2, the comparison operations for 128-bit modular arithmetic require more instructions and additional handling compared to AVX-512.

\bparagraph{From arithmetic to cryptographic kernels.}
At this point, we have discussed how to build efficient scalar and SIMD-vectorized double-word modular arithmetic. Extending modular arithmetic to BLAS operations is relatively straightforward, as BLAS operations are essentially vector-based modular arithmetic (e.g., a vector of 1,024 elements, not a SIMD vector). They can be implemented by looping over scalar or SIMD modular arithmetic. In cryptographic settings, the vector length used in BLAS operations is typically a power of two, and in this work, we assume it is a multiple of the SIMD lane size we use. 
Transitioning from double-word modular arithmetic to NTTs is more complicated. On the computation side, the core NTT building block is the \textit{butterfly}, which consists of one modular addition, one modular subtraction, and one modular multiplication. An $n$-point NTT consists of $\log n$ stages, with each stage containing $n/2$ butterflies. The results from each butterfly in the previous stage need to be communicated to the next stage. We build upon prior work that implements AVX-512-based NTT~\cite{fu2024avxntt}, which employs the Pease algorithm~\cite{pease1968adaptation} for the NTT dataflow. The data permutation stage requires the use of the unpack and permute instructions, such as  \verb|_mm512_unpacklo_epi64|, \verb|_mm512_unpackhi_epi64|, and \verb|_mm512_permutex2var_epi64| in AVX-512.

\section{Multi-word Extension (MQX)}

Although SIMD vectorization enhances the performance of the scalar implementation, it still falls short of the levels achievable on ASICs based on our empirical results (discussed in Section~\ref{sec:exp}). 
We compare the scalar implementation of double-word addition with carry against the AVX-512 implementation in Table~\ref{tab:adc} and observe that the scalar code can be captured in x86 assembly as a single ALU instruction, \verb|ADC|, whereas the AVX-512 implementation requires six SIMD instructions to achieve the same functionality.
Therefore, to address the performance bottlenecks of existing SIMD instructions in cryptographic settings, we propose multi-word extension (MQX), an ISA extension designed to support the large integer arithmetic used by cryptographic kernels while leveraging SIMD vector parallelism. 
The design philosophy behind MQX is to introduce minimal changes to the existing ISA. 
We expect the required engineering effort to be manageable, given MQX's close similarity to prior vector instruction sets, and we introduce our performance modeling of MQX based on existing AVX-512 instructions.
Table~\ref{tab:mqx} presents the proposed MQX instruction set that extends AVX-512 (referred to as AVX-512 MQX). However, MQX is not limited to AVX-512; it can also be integrated with other SIMD instruction sets, such as AVX2 and Neon, since both the word size and the number of SIMD lanes are configurable in MQX. In this work, we focus on AVX-512 MQX and refer to it simply as MQX. 

\begin{table*}[!htb]
\small
    \caption{AVX-512 multi-word extension (MQX).}
    \centering
    \label{tab:mqx}
    \begin{tabular}{p{0.33\linewidth} p{0.28\linewidth} p{0.3\linewidth}}
    \toprule[1pt]
    \begin{normalsize}\textbf{Instruction}\end{normalsize} & \begin{normalsize}\textbf{Emulation}\end{normalsize} & \begin{normalsize}\textbf{Description}\end{normalsize} \\
    \midrule
\vspace{-3.5mm}
\begin{lstlisting}[columns=fullflexible, numbers=none]
void _mm512_mul_epi64(__m512i* ch, __m512i* cl,
     __m512i a, __m512i b)
\end{lstlisting}
& 
\vspace{-3.5mm}
\begin{lstlisting}[columns=fullflexible, numbers=none]
for (i = 0; i < 8; i++) {
    *ch[i] = ((i128) a[i] * (i128) b[i]) 
             >> 64;
    *cl[i] = a[i] * b[i]; }
\end{lstlisting}
&
Multiply two words, and output the high and low parts of the result as two words.
    \\
\vspace{-5.5mm}
\begin{lstlisting}[columns=fullflexible, numbers=none]
__m512i _mm512_adc_epi64(__m512i a, __m512i b,
        __mmask8 ci, __mmask8* co)
\end{lstlisting}
& 
\vspace{-5.5mm}
\begin{lstlisting}[columns=fullflexible, numbers=none]
for (i = 0; i < 8; i++) {
    *co[i] = ((i128) a[i] + (i128) b[i] 
             + (i128) ci[i]) >> 64;
    c[i] = a[i] + b[i] + ci[i]; 
    return c; }
\end{lstlisting}
& 
\vspace{-5.5mm}
Add two words and a carry bit, and output a word and a carry bit. 
    \\
\begin{minipage}[t]{\linewidth}
\vspace{-5.5mm}
\begin{lstlisting}[columns=fullflexible, numbers=none]
__m512i _mm512_sbb_epi64(__m512i a, __m512i b, 
        __mmask8 bi, __mmask8* bo)
\end{lstlisting}
\end{minipage}
& 

\begin{minipage}[t]{\linewidth}
\vspace{-5.5mm}
\begin{lstlisting}[columns=fullflexible, numbers=none]
for (i = 0; i < 8; i++) {
    *bo[i] = ((i128) a[i] - (i128) b[i] 
        - (i128) bi[i]) >> 127;
    c[i] = a[i] - b[i] - bi[i]; 
    return c; }
\end{lstlisting}
\end{minipage}
\vspace{0.2mm}
& 
\vspace{-5.5mm}
Subtract two words and a borrow bit, and output a word and a borrow bit.
    \\ 
    \bottomrule[1pt]
    \end{tabular}
\small
\end{table*}

\subsection{Design Philosophy}

The design philosophy of MQX is to minimize the required engineering effort from the hardware side. To this end, we propose only three new instructions, all in SIMD form: i) widening multiplication, ii) addition with carry, and iii) subtraction with borrow. Each instruction is based on features already present in the x86 architecture. 
We also find strong similarities between the proposed MQX instructions and the SIMD counterparts implemented by Intel in the Larrabee architecture~\cite{seiler2008larrabee}. 
Introduced around 2008, Larrabee featured a new ISA known as Larrabee New Instructions (LRBni), which extended in-order x86 CPU cores with wide vector processing units and fixed-function logic blocks. Its goal was to achieve higher performance per watt and per unit area than out-of-order CPUs, particularly for highly parallel workloads.
The Larrabee architecture later evolved into the Intel Many Integrated Core (MIC) architecture, with the first commercial product released as the Knights Corner (KNC) processor, whose intrinsics were documented in the Intel Intrinsics Guide~\cite{intel2013guide} from versions 3.1 to 3.6.5.
Built upon the foundations of both x86 and previously proposed SIMD extensions, each MQX instruction will be introduced in detail in the following paragraphs.

\begin{lstlisting}[columns=fullflexible, captionpos=b, float, floatplacement=t, caption={Double-word modular addition using MQX.}, label={lst:dadd_mqx}]
    void addmod128(__m512i* ch, __m512i* cl, __m512i ah, __m512i al, 
                   __m512i bh, __m512i bl, __m512i mh, __m512i ml) {
        __m512i el, eh, c1; __mmask8 elc, ehc, ehc1, ctrl, clc;
        el = (*@\textcolor[HTML]{B22222}{\_mm512\_adc\_epi64(al, bl, z\_mask, \&elc);}@*)
        eh = (*@\textcolor[HTML]{B22222}{\_mm512\_adc\_epi64(ah, bh, elc, \&ehc);}@*)
        ehc1 = _mm512_cmp_epi64_mask(mh, eh, _MM_CMPINT_LT);
        ctrl = (ehc1 | ehc);
        c1 = (*@\textcolor[HTML]{B22222}{\_mm512\_sbb\_epi64(el, ml, z\_mask, \&clc);}@*)
        *cl = _mm512_mask_blend_epi64(ctrl, el, c1);
        c1 = (*@\textcolor[HTML]{B22222}{\_mm512\_sbb\_epi64(eh, mh, clc, \&ehc);}@*)
        *ch = _mm512_mask_blend_epi64(ctrl, eh, c1);
    }
\end{lstlisting}

\bparagraph{Widening multiplication.} \begin{footnotesize}\verb|void _mm512_mul_epi64(__m512i* ch|, \verb|__m512i* cl, __m512i a, __m512i b)|\end{footnotesize} is a widening multiplication. For each SIMD lane, it multiplies two 64-bit words and stores the high part of the result in one 64-bit word and the low part in another. This mirrors the behavior of the unsigned multiply \verb|MUL| in x86, where the result is stored in a register pair containing both high and low parts. In AVX-512, there is support for variants of lower bit-width such as \begin{footnotesize}\verb|__m512i _mm512_mul_epi32 (__m512i a, __m512i b)|\end{footnotesize}, which multiplies the low signed 32-bit integers in each 64-bit SIMD lane and stores the 64-bit result. However, for 64-bit inputs in AVX-512, we only have multiply-low as \begin{footnotesize}\verb|__m512i _mm512_mullo_epi64 (__m512i a, __m512i b)|\end{footnotesize} that returns only the lower 64 bits of the result.
In LRBni, Intel introduced a vector multiply-high \verb|vmulhpi| for 32-bit signed integers. Similarly, the KNC architecture includes \begin{footnotesize}\verb|__m512i _mm512_mulhi_epi32 (__m512i a, __m512i b)|\end{footnotesize}, as documented in earlier versions of the Intel Intrinsics Guide.
In Section~\ref{sec:sa}, we evaluate whether the hardware engineering cost of a widening multiply can be justified and analyze the performance trade-offs of implementing only multiply-high instead of a full widening multiplication. 

\bparagraph{Addition with carry.} 
\begin{footnotesize}\verb|__m512i _mm512_adc_epi64(__m512i a,| \verb|__m512i b, __mmask8 ci, __mmask8* co)|\end{footnotesize} performs a per-lane 64-bit addition with carry-in and outputs both the addition result and carry-out. This directly mirrors the behavior of \verb|ADC| in x86, where the carry flag \verb|CF| is used as input and is also updated as part of the result. 
In LRBni, \verb|vadcpi| performs vector addition with carry-in and carry-out on 32-bit integers. There exists a 32-bit counterpart in KNC as well: \begin{footnotesize}\verb|__m512i _mm512_adc_epi32 (__m512i v2, __mmask16 k2, __m512i v3,| \verb|__mmask16* k2_res)|\end{footnotesize}.

\bparagraph{Subtraction with borrow.} 
Similar to addition with carry, \begin{footnotesize}\verb|__m512i _mm512_sbb_epi64(__m512i a, __m512i b, __mmask8 bi, __mmask8 bo)|\end{footnotesize} implements a per-lane 64-bit subtraction with borrow-in and outputs the subtraction result with borrow-out. This instruction mirrors the behaviors of \verb|SBB| in x86, where the carry flag \verb|CF| (used for borrow) is taken as input and updated based on the result.
In LRBni, the corresponding instruction is \verb|vsbbpi|, which performs vector subtraction with borrow-in and borrow-out for 32-bit integers. Similarly, in KNC, we can find the 32-bit version of subtraction with borrow as \begin{footnotesize}\verb|__m512i _mm512_sbb_epi32 (__m512i v2, __mmask16 k, __m512i v3,| \verb|__mmask16* borrow)|\end{footnotesize}.

The C code for double-word modular addition using the proposed MQX instructions is provided in Listing~\ref{lst:dadd_mqx}, where \verb|z_mask| is a global 8-bit mask of zeros. 
In summary, MQX is designed with the goal of requiring relatively low hardware engineering effort. Each MQX instruction has a scalar equivalent and a closely related vector counterpart that has previously been proposed or implemented by Intel. The core idea behind MQX is to introduce wider multiplication units and first-class support for carry and borrow in vector operations, enabling more efficient handling of large integer arithmetic for cryptographic kernels.

\subsection{Performance Projection using Proxy ISA (PISA)}
\label{sec:pisa}

After proposing the MQX instructions, we aim to evaluate the performance gains they offer and assess whether these gains justify implementing the instructions in hardware.

While open-source simulation tools such as gem5~\cite{binkert2011gem5} and zSim~\cite{sanchez2013zsim} offer detailed evaluations of microarchitectural designs, they fall short in capturing the proprietary designs developed by manufacturers like Intel and AMD. In particular, zSim was primarily designed for Haswell-era processors and has not been actively updated to support newer architectures: The majority of its codebase was committed over a decade ago, with the latest update occurring two years ago. The persistent lack of support for modern microarchitectures stems from the significant engineering effort required to build accurate simulators, compounded by the proprietary and undocumented nature of commercial CPU designs.
Meanwhile, gem5 offers only limited support for native AVX intrinsics~\cite{lee2024gem5}. In principle, both baseline AVX and MQX could be implemented in gem5, but doing so would require substantial engineering effort---particularly given that the last officially published gem5 extension only partially implemented the SSE family that dates back over two decades. There have been two open-source, community-driven efforts to extend gem5 with AVX-512 support; however, both remain incomplete. Zhang's work~\cite{gem5avx2022} lacks masking support (which is essential for modular arithmetic), while Lee et al.'s work~\cite{lee2024gem5} omits memory alignment support.

To facilitate the performance modeling process, we propose a new performance modeling approach named performance projection using proxy ISA (PISA). PISA allows us to estimate the performance of MQX without trying to reproduce or reverse engineer the proprietary microarchitectures from Intel or AMD. Specifically, we designate AVX-512 as the proxy ISA and project the performance of MQX by mapping each MQX instruction to the most structurally similar AVX-512 instruction. In other words, if implemented efficiently, each MQX instruction could achieve performance comparable to its AVX-512 proxy. The specific proxy instructions in AVX-512 used in our modeling are shown in Table~\ref{tab:mqx_pisa}.

\begin{table}[!htb]
\small
    \caption{Proxy instructions in AVX-512 for MQX performance projection.}
    \centering
    \rowcolors{2}{white}{gray!15}
    \label{tab:mqx_pisa}
    \begin{tabular}{p{0.45\linewidth} p{0.45\linewidth}}
    \toprule[1pt]
    \begin{normalsize}\textbf{MQX instruction}\end{normalsize} & \begin{normalsize}\textbf{AVX-512 proxy instruction}\end{normalsize} \\
    \midrule
    \verb|_mm512_mul_epi64| & \verb|_mm512_mullo_epi64| \\
    \verb|_mm512_adc_epi64| & \verb|_mm512_mask_add_epi64| \\
    \verb|_mm512_sbb_epi64| & \verb|_mm512_mask_sub_epi64| \\
    \bottomrule[1pt]
    \end{tabular}
\small
\end{table}

PISA enables us to receive fast and early feedback in the software-hardware co-design cycle. By modeling many potential instructions via proxy mappings, PISA enables quick exploration of different ISA extensions. Moreover, rather than relying on simulation results, PISA allows us to report real measurements on actual hardware, grounding our projections in real-world performance.
We view PISA and microarchitectural simulators as complementary: PISA offers a top-down, actionable alternative for performance modeling and helps assess whether an ISA extension is promising enough to warrant the substantial engineering effort required to implement it in cycle-level simulators or in other tools for detailed analysis, such as power and area estimation.

We are extremely careful in choosing proxy instructions. A key observation that grounds our performance projections is that, based on both Intel’s optimization manual~\cite{optm2023intel} and empirical data\footnote{\url{https://uops.info/table.html}}, the x86 instruction set shows no measurable latency difference between \verb|ADD| and \verb|ADC|, between \verb|SUB| and \verb|SBB|, or between 32-bit and 64-bit \verb|MUL| on both Ice Lake and Zen 4 architectures, which we used to benchmark MQX's performance in Section~\ref{sec:exp}.
While we acknowledge that extending scalar performance behavior to SIMD  is nontrivial, these observations provide a reasonable foundation for our performance projections. Furthermore, in Section~\ref{sec:validating_pisa}, we present sanity checks that validate the proposed PISA methodology.

\begin{figure}[t]
\centering
    \includegraphics[width=0.99\columnwidth,trim={66mm 40mm 110mm 50mm},clip]{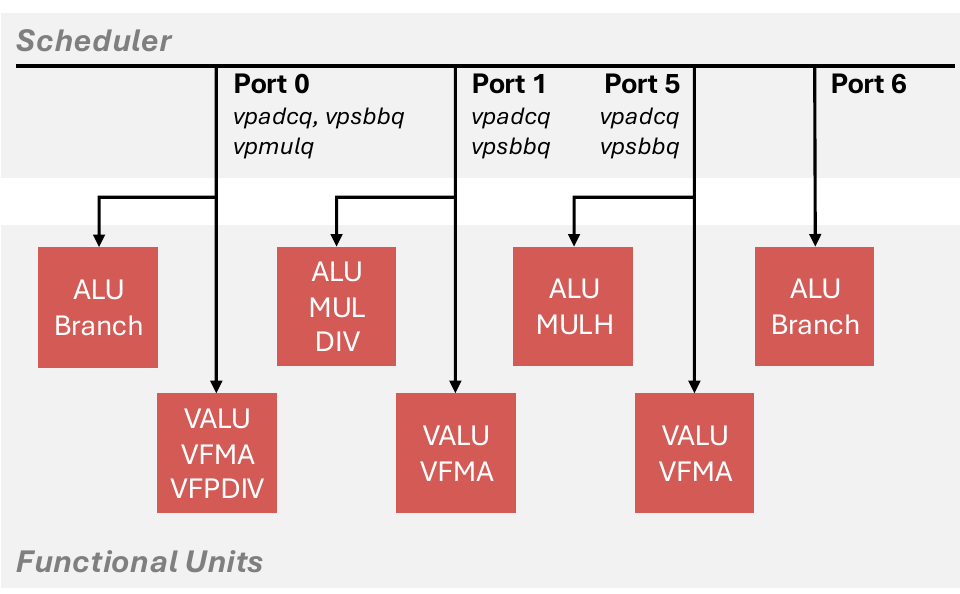}
    \caption{Simplified Sunny Cove microarchitecture used by Intel Xeon.}
    \label{fig:intel_march}
\end{figure}

\bparagraph{Machine code analysis.}
Using PISA, we are able to leverage machine code analysis tools such as LLVM-MCA to simulate how each MQX instruction would be scheduled on CPUs, assuming each instruction maps to the same execution port as its proxy ISA counterpart. We apply LLVM-MCA to analyze the behavior of modular addition, subtraction, and multiplication with AVX-512 and MQX on Intel Xeon 8352Y, which is based on the Sunny Cove microarchitecture (a simplified version shown in Figure~\ref{fig:intel_march}).

Listing~\ref{lst:dadd_mca} shows the LLVM-MCA analysis of double-word modular addition with AVX-512 and MQX. To construct the MQX part, we start from the LLVM-MCA analysis of the code used to estimate MQX performance with PISA and then manually edit the original AVX-512 assembly into its MQX counterpart (e.g., \verb|vpsubq| is replaced by \verb|vpsbbq|, and a mask register \verb|k0| is prepended, which \verb|vpsbbq| writes to). This modified listing is \textit{not} semantically correct; rather, it serves as a performance-oriented demonstration that illustrates how MQX instructions would appear in assembly and how they would map to the same execution ports as their AVX-512 proxies.

\begin{lstlisting}[columns=fullflexible, numbers=none, basicstyle=\scriptsize\ttfamily, captionpos=b, float, floatplacement=t, caption={LLVM-MCA analysis (edited for brevity) of double-word modular addition with AVX-512 and MQX. The MQX instructions, shown in red, are included as a performance-oriented illustration. Square brackets \texttt{[}\#\texttt{]} denote port numbers.}, label={lst:dadd_mca}]
AVX-512 - Resource pressure by instruction:
[0]    [1]    [2]    [3]    [4]    [5]    Instructions:
 -     1.00    -      -      -      -     vpaddq     %zmm2, %zmm3, %zmm3
 -      -      -      -      -     1.00   vpcmpltuq  %zmm2, %zmm3, %k1
 -     1.00    -      -      -      -     vpaddq     %zmm4, %zmm6, %zmm2
 -     1.00    -     1.00    -      -     vmovdqu64  one(%rip), %zmm5
1.00    -      -      -      -      -     vpaddq	 %zmm5, %zmm2, %zmm2 {%k1}
 -      -      -      -      -     1.00   vpcmpnleuq %zmm0, %zmm2, %k0
 -      -      -      -      -     1.00   vpcmpeqq   %zmm0, %zmm2, %k1
 -      -      -      -      -     1.00   vpcmpltuq  %zmm1, %zmm3, %k2
 -      -      -      -      -     1.00   vpcmpnltuq %zmm1, %zmm3, %k1 {%k1}
 -      -      -      -      -     1.00   vpmaxuq    %zmm6, %zmm4, %zmm4
 -      -      -      -      -     1.00   vpcmpnleuq %zmm2, %zmm4, %k3
1.00    -      -      -      -      -     korb	     %k0, %k3, %k0
1.00    -      -      -      -      -     korb	     %k1, %k0, %k1
1.00    -      -      -      -      -     vmovdqa64	 %zmm0, %zmm4
 -     1.00    -      -      -      -     vpaddq     %zmm5, %zmm0, %zmm4 {%k2}
 -     1.00    -      -      -      -     vpsubq     %zmm4, %zmm2, %zmm2 {%k1}
 -      -      -      -      -     1.00   vpsubq     %zmm1, %zmm3, %zmm3 {%k1}

MQX - Resource pressure by instruction:
[0]    [1]    [2]    [3]    [4]    [5]    Instructions:
 -     1.00    -      -      -      -     (*@\textcolor[HTML]{B22222}{vpadcq  \%zmm7, \%zmm3, \%zmm3 \{\%k1\} \{\%k3\}}@*)
 -     1.00    -      -      -      -     vmovq  %xmm6, %xmm4
1.00    -      -      -      -      -     (*@\textcolor[HTML]{B22222}{vpadcq  \%zmm5, \%zmm4, \%zmm4 \{\%k3\} \{\%k0\}}@*)
 -      -      -      -      -     1.00   vpcmpgtq %zmm0, %zmm4, %k2
1.00    -      -      -      -      -     korb   %k0, %k2, %k2
 -      -      -      -      -     1.00   (*@\textcolor[HTML]{B22222}{vpsbbq  \%zmm1, \%zmm3, \%zmm3 \{\%k1\} \{\%k3\}}@*)
 -     1.00    -      -      -      -     (*@\textcolor[HTML]{B22222}{vpsbbq	\%zmm2, \%zmm4, \%zmm4 \{\%k3\} \{\%k0\}}@*)
\end{lstlisting}

\bparagraph{Functional correctness.} In our MQX implementation, we have a flag to check for functional correctness. When the flag is turned off, we execute the code using PISA with the expectation of not getting correct results. With that flag turned on, each MQX instruction is emulated by a scalar implementation as shown in Table~\ref{tab:mqx}. Hence, we can obtain functional correctness of MQX and verify that we are not missing any instructions. With PISA, we also carefully inspect the compiler-generated assembly code to make sure no instructions are incorrectly pruned.

\section{Evaluation}
\label{sec:exp}

In this section, we describe our experimental setup and discuss our results of scalar, SIMD-vectorized, and MQX-based cryptographic kernels. We also conduct two sets of sensitivity analyses: one on the MQX instructions and another on the choice between two multiplication algorithms.

\subsection{Experimental Setup}

We evaluated our proposed approaches on two state-of-the-art server-grade CPUs: Intel Xeon 8352Y, provided as part of the FASTER node at Texas A\&M University, and AMD EPYC 9654, provided as part of the Launch node at the same institution. Detailed specifications for each CPU are shown in Table~\ref{tab:cpus}. For brevity, we refer to Intel Xeon 8352Y as Intel Xeon and AMD EPYC 9654 as AMD EPYC in the remainder of this paper.

\begin{figure*}[t]
\centering
    \subfloat[Intel Xeon.\label{fig:blas_intel}]{\centering\includegraphics[width=\columnwidth,trim={8mm 5mm 12mm 3mm},clip]{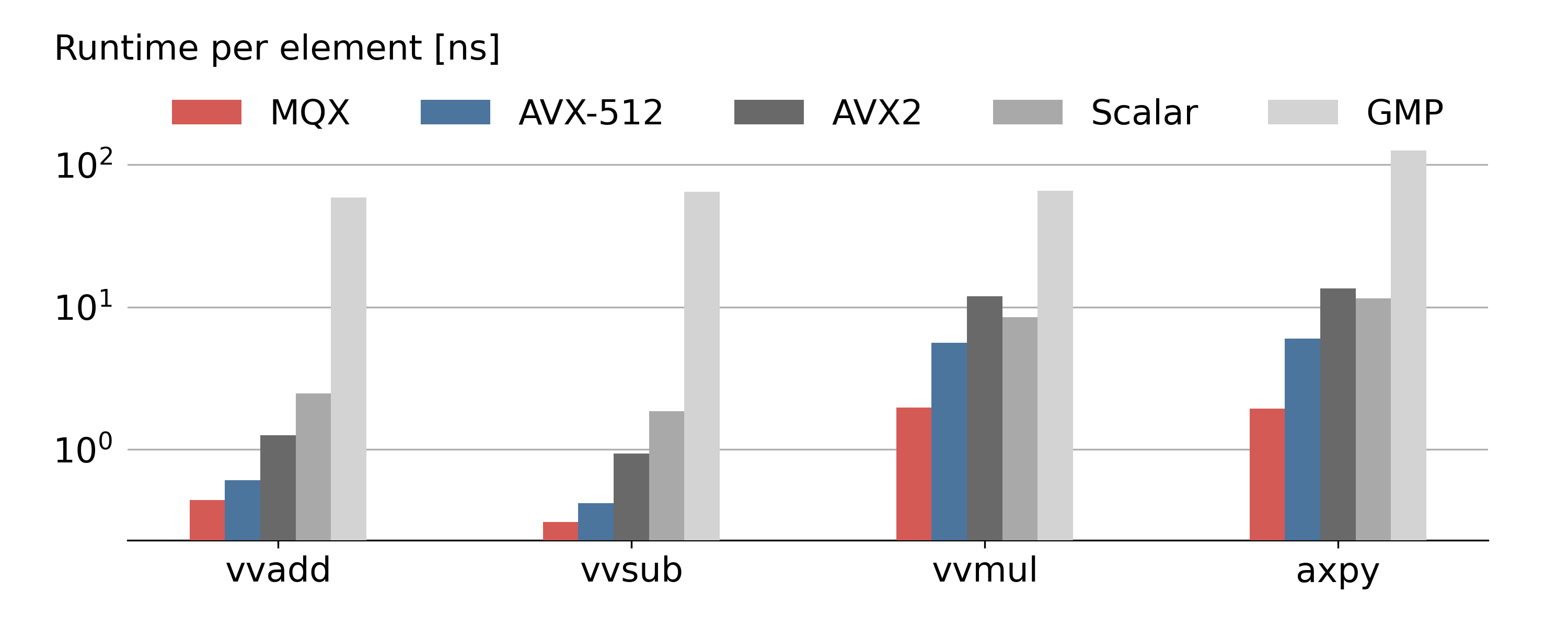}} 
    \hspace{6mm}
    \subfloat[AMD EPYC.\label{fig:blas_amd}]{\centering\includegraphics[width=\columnwidth,trim={8mm 5mm 12mm 3mm},clip]{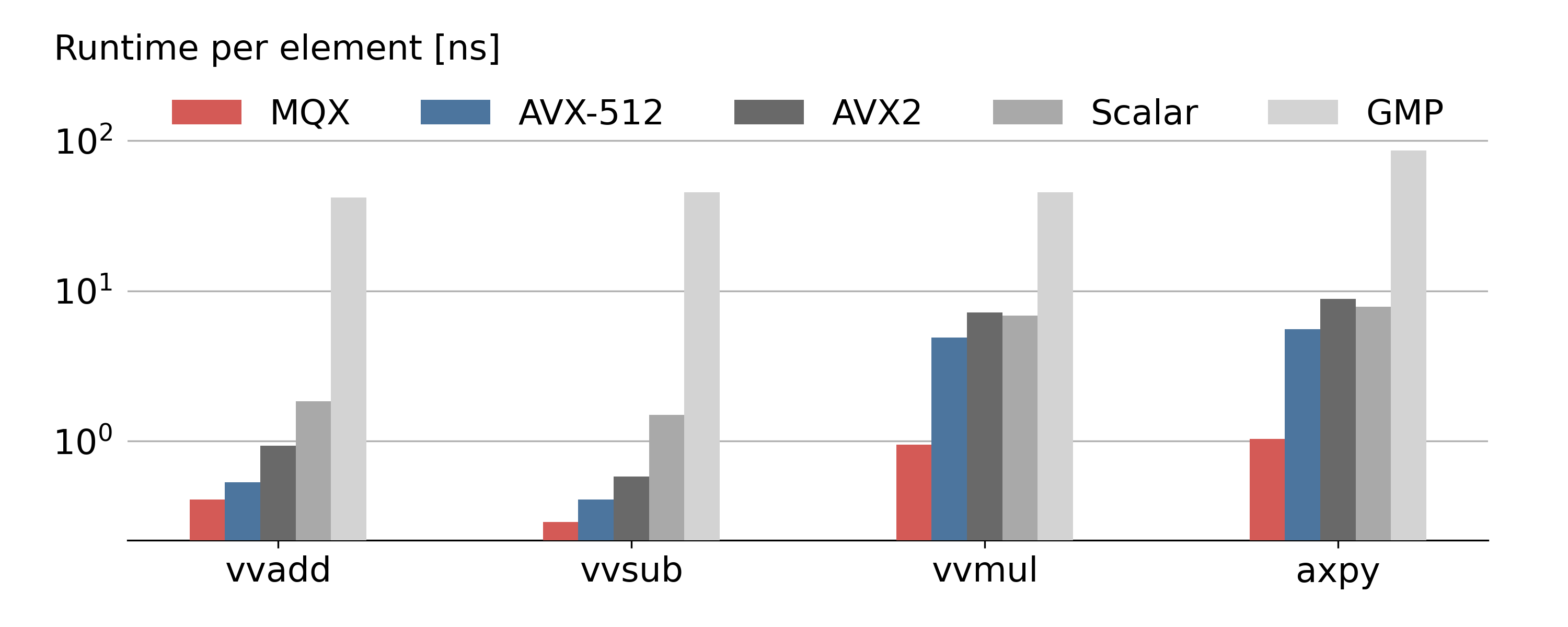}}
    \caption{Performance of BLAS operations on a single CPU core.}
    \vspace{-3mm}
\end{figure*}

\begin{table}[!htb]
    \caption{CPUs used for benchmarking.}
    \centering
    \rowcolors{2}{white}{gray!15}
    \label{tab:cpus}
    \begin{tabular}{>{\raggedright\arraybackslash}p{0.28\linewidth} 
        >{\raggedleft\arraybackslash}p{0.29\linewidth} 
        >{\raggedleft\arraybackslash}p{0.29\linewidth}}
    \toprule[1pt]
    \textbf{Specification} & \textbf{Intel Xeon 8352Y}  & \textbf{AMD EPYC 9654} \\
    \midrule
    \textbf{Base Clock Rate} & 2.2 GHz           & 2.4 GHz       \\
    \textbf{Max Clock Rate}  & 3.4 GHz           & 3.7 GHz       \\
    \textbf{Memory}          & 256 GB DDR4       & 384 GB DDR5   \\     
    \textbf{L3 Cache}        & 48 MB             & 384 MB        \\     
    \bottomrule[1pt]
    \end{tabular}
\end{table}

\bparagraph{Measuring kernel runtime.}
We use the Intel oneAPI DPC++/C++ compiler (ICX, version 2024.2.0) to compile all kernels on Intel Xeon, and the AMD Optimizing C/C++ and Fortran Compiler (AOCC, version 4.1.0) to compile all kernels on AMD EPYC. 
For NTTs, we report the average runtime of the final 50 iterations out of 100 runs; for BLAS operations, we report the average runtime of the final 500 iterations out of 1,000 runs. This approach allows the cache to warm up and stabilize, and mitigates fluctuations in measured runtime (for BLAS operations, in particular). 
For all BLAS operations, the vector length is set to 1,024, as it aligns with typical polynomial sizes in FHE schemes. 
Although we have implemented both the Karatsuba and the schoolbook algorithm for modular multiplication, we only report the runtime of the schoolbook implementation, as it consistently outperforms Karatsuba (see Section~\ref{sec:sa} for further discussion).
Our timing measurements include data transfer time.

\subsection{Validating PISA}
\label{sec:validating_pisa}

Before presenting the performance results for BLAS operations and NTTs, we first perform a sanity check on our PISA performance modeling methodology, which underpins our analysis of MQX.
Specifically, to validate PISA, we applied the same modeling methodology used for MQX to \textit{existing} SIMD instructions used in our NTT implementations so that we could establish ground truth and assess the accuracy of PISA's performance predictions.

As shown in Table~\ref{tab:validating_pisa}, for multiplication, we use the existing AVX2 widening multiplication on 32-bit elements, \verb|_mm256_mul_epu32|, as the ground truth and \verb|_mm256_mullo_epi32| as the proxy instruction (since there is no \verb|_mm256_mullo_epu32|). This closely mirrors how we model the widening multiplication instruction proposed in MQX, as shown in Table~\ref{tab:mqx_pisa}. Note that \verb|_mm512_mul_epu32| in AVX-512 is not used as the ground truth, even though it more closely resembles our proposed instruction. The reason is that our AVX-512 NTT implementation operates on 64-bit aligned data and does not utilize 32-bit widening multiplication, and adapting 32-bit SIMD instructions would require re-implementing the entire NTT kernel. For addition and subtraction, we use regular SIMD addition to model masked addition in AVX-512. Consistent with our conservative modeling methodology for MQX, we insert an extra instruction and guard the output with \verb|volatile| to preserve data dependencies on the mask register. 
This set of experiments is designed to demonstrate PISA's capability of handling cases where the proxy instruction does not take identical input arguments as the target instruction.

\begin{table}[!htb]
\small
    \caption{Target and proxy instructions for validating PISA.}
    \centering
    \rowcolors{2}{white}{gray!15}
    \label{tab:validating_pisa}
    \begin{tabular}{p{0.45\linewidth} p{0.45\linewidth}}
    \toprule[1pt]
    \begin{normalsize}\textbf{Target instruction}\end{normalsize} & \begin{normalsize}\textbf{Proxy instruction}\end{normalsize} \\
    \midrule
    \verb|_mm256_mul_epu32| & \verb|_mm256_mullo_epi32| \\
    \verb|_mm512_mask_add_epi64| & \verb|_mm512_add_epi64| \\
    \verb|_mm512_mask_sub_epi64| & \verb|_mm512_sub_epi64| \\
    \bottomrule[1pt]
    \end{tabular}
\small
\end{table}

To evaluate the accuracy of the projected performance, we define the relative error $\varepsilon$ as
\begin{equation}
\varepsilon = \frac{t_{\text{target}} - t_{\text{proxy}}}{t_{\text{target}}} \cdot 100\%,
\end{equation}
where $t_\text{target}$ denotes the NTT runtime using the target instruction, and $t_\text{proxy}$ denotes the runtime using the proxy instruction.

We validated PISA using an NTT size of $2^{14}$, the average among the NTT sizes targeted in this paper. As shown in Table~\ref{tab:pisa_results}, the absolute relative error of using PISA is below 8\% for all six test cases. Negative values indicate that PISA is overly conservative, projecting higher runtimes than observed. In all cases where PISA is overly optimistic, the relative error remains under 6\%. The results suggest that PISA passes a sanity check when modeling both AVX2 and AVX-512 instructions on the Intel and AMD CPUs we tested.

\begin{table}[!htb]
\small
    \caption{Relative error ($\boldsymbol{\varepsilon}$) of PISA-projected runtime on Intel and AMD CPUs.}
    \centering
    \rowcolors{2}{white}{gray!15}
    \label{tab:pisa_results}
    \begin{tabular}{>{\raggedright\arraybackslash}p{0.42\linewidth} 
        >{\raggedleft\arraybackslash}p{0.22\linewidth} 
        >{\raggedleft\arraybackslash}p{0.22\linewidth}}
    \toprule[1pt]
    \begin{normalsize}\textbf{Target instruction}\end{normalsize} & \begin{normalsize}\textbf{Intel Xeon}\end{normalsize} & \begin{normalsize}\textbf{AMD EPYC}\end{normalsize} \\
    \midrule
    \verb|_mm256_mul_epu32| & $3.23$\% & $2.64$\% \\
    \verb|_mm512_mask_add_epi64| & $-7.68$\% & 5.25\% \\
    \verb|_mm512_mask_sub_epi64| & $-4.30$\% & 1.27\% \\
    \bottomrule[1pt]
    \end{tabular}
\small
\end{table}

\vspace{-3mm}
\subsection{BLAS Operation Results}
\label{sec:results_blas}

\begin{figure*}[t]
\centering
    \subfloat[Intel Xeon.\label{fig:ntt_intel}]
    {\centering\includegraphics[width=\columnwidth,trim={20mm 3mm 20mm 5mm},clip]{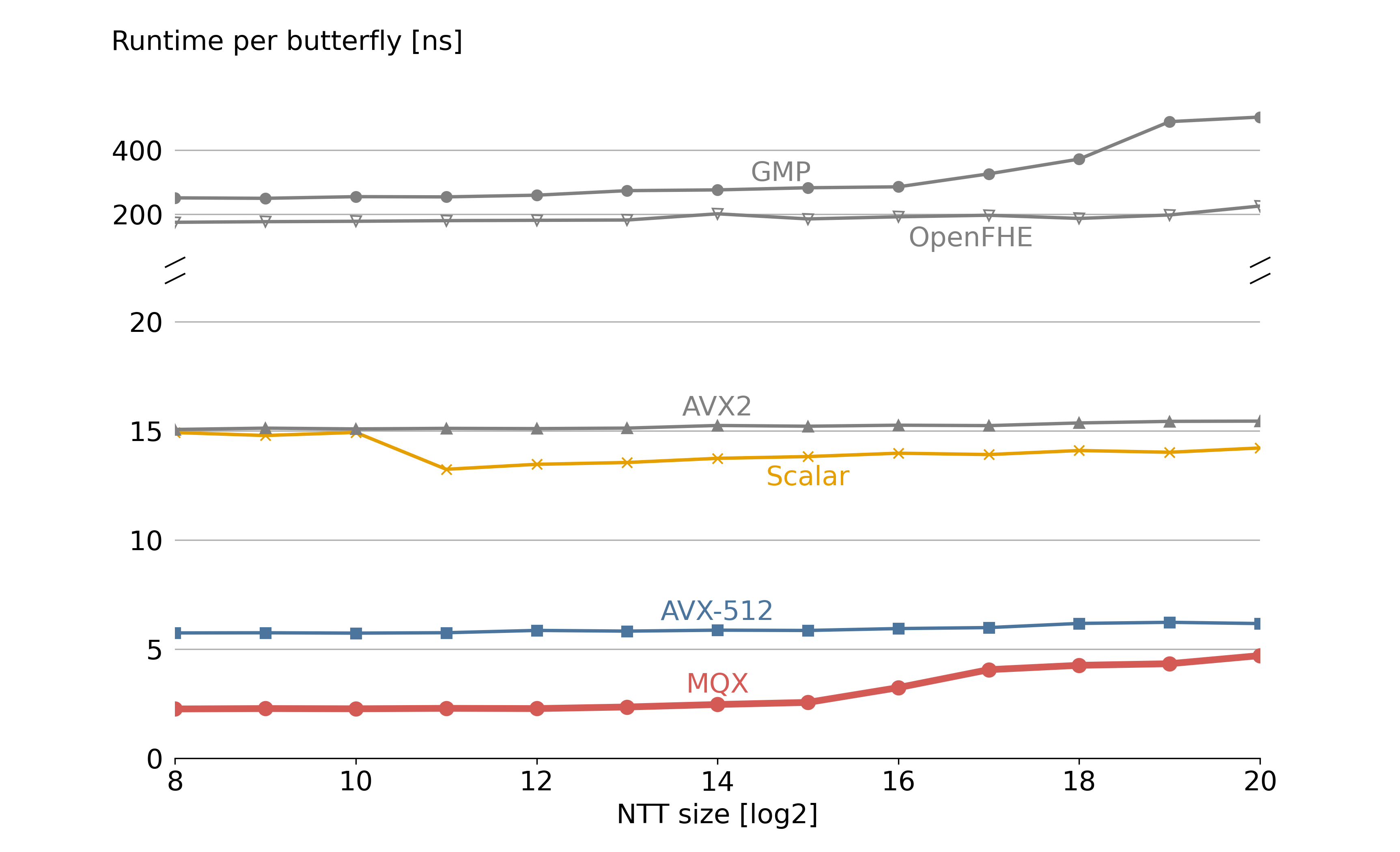}}
    \hspace{6mm}
    \subfloat[AMD EPYC.\label{fig:ntt_amd}]{\centering\includegraphics[width=\columnwidth,trim={20mm 3mm 20mm 5mm},clip]{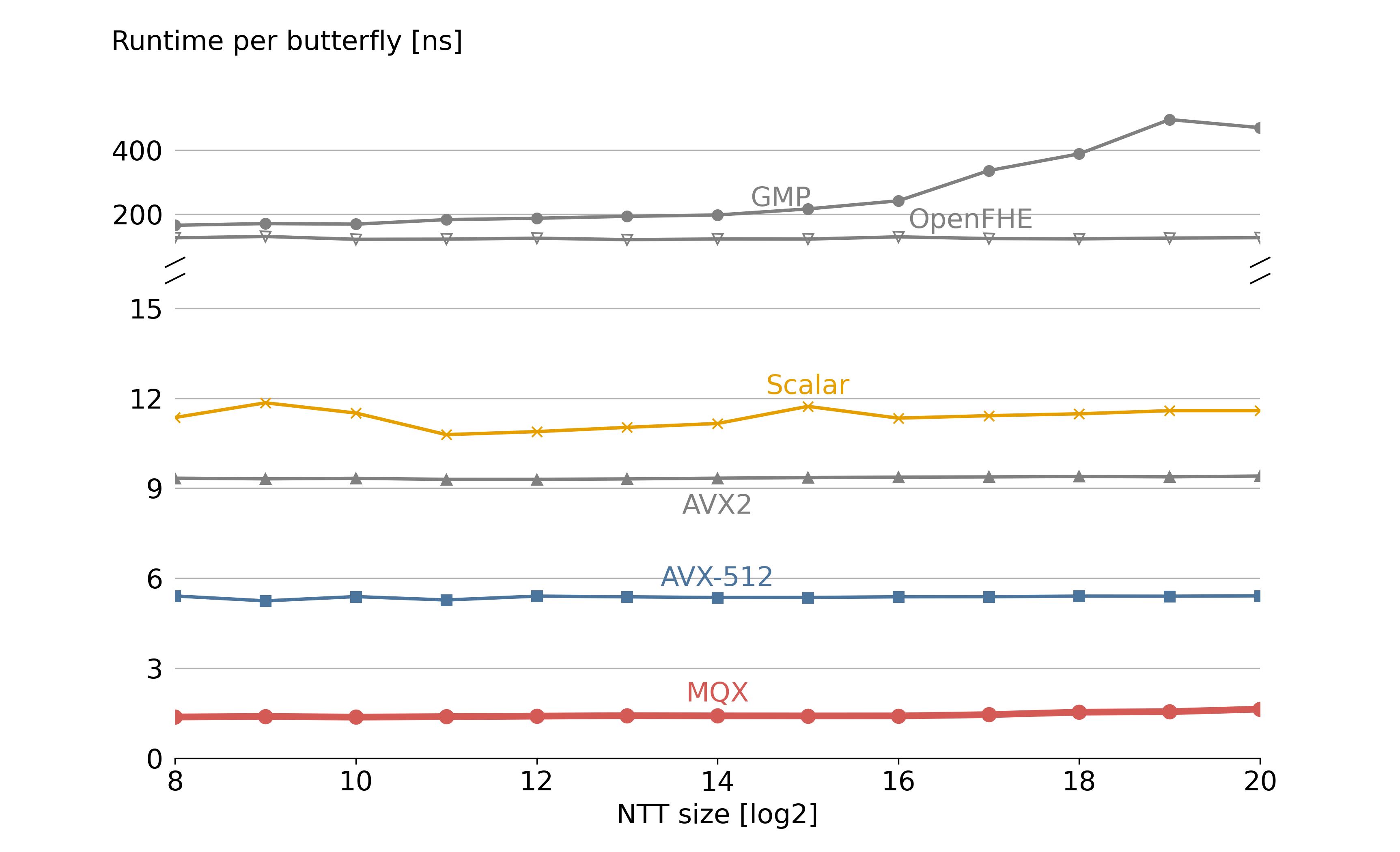}} \\ 
    \caption{Performance of NTT on a single CPU core.}
    \vspace{-3mm}
\end{figure*}

We evaluated the performance of our BLAS operation implementations using scalar, AVX2, AVX-512, and MQX, and compared them against a baseline implementation using the state-of-the-art arbitrary-precision library GMP. The GMP library was configured to perform exact integer arithmetic, ensuring bitwise-identical results with both our implementation and other baselines.
Figure~\ref{fig:blas_intel} shows the results on Intel Xeon. On Intel Xeon, the scalar implementation of NTT outperforms AVX2 in vector multiplication and axpy, but falls behind in vector addition and subtraction. AVX-512 delivers an average speedup of 2.2 times over AVX2 across four BLAS operations. MQX further improves performance, achieving a 2.2 times speedup over AVX-512 and outperforming all other baselines. GMP exhibits a slowdown of 18.4 times compared to the slower of the AVX2 and scalar implementations.

As shown in Figure~\ref{fig:blas_amd}, we observe a similar performance trend on AMD EPYC as on Intel Xeon: MQX performs best, followed by AVX-512, AVX2, scalar, and GMP. AVX2 performs slightly worse than the scalar implementation in vector multiplication and axpy, but achieves a 1.6 times overall speedup across all four BLAS operations. On AMD EPYC, the performance gap between AVX2 and AVX-512 is smaller than on Intel Xeon, with AVX2 falling behind AVX-512 by 1.6 times on average. MQX delivers a 3.2 times speedup over AVX-512, while GMP shows a 17.3 times slowdown compared to the slowest of our implementations.

In summary, by leveraging natively supported SIMD instructions (i.e., AVX-512), we achieve an average speedup of 62 times over the state-of-the-art arbitrary-precision library across the two CPUs evaluated. With the proposed MQX ISA extension, we gain an additional average speedup of 2.7 times over AVX-512.

\subsection{NTT Results}
\label{sec:ntt_results}

For the 128-bit bit-width targeted in this work, we compared against two state-of-the-art libraries. The first one is GMP~\cite{granlund1996gnu}, a general-purpose arbitrary-precision library. 
The second baseline is the open-source fully homomorphic encryption library (OpenFHE)~\cite{al2022openfhe}, one of the most widely used open-source FHE libraries that offers efficient implementations of all major FHE schemes. In our experiments, we used its default mathematical backend for large integer arithmetic.
We evaluated the performance of our proposed NTT implementations against these two baselines, with all experiments conducted on a single CPU core.

\bparagraph{Intel Xeon.} As shown in Figure~\ref{fig:ntt_intel}, on Intel Xeon, our scalar implementation outperforms the state-of-the-art FHE library OpenFHE~\cite{al2022openfhe} by 13.5 times on average across all NTT sizes. 
Notably, AVX2 and scalar implementations achieve comparable performance, with the scalar implementation being slightly faster. Using AVX-512 yields a 2.4 times speedup over the scalar implementation, increasing the performance gap against the best baseline to 31.9 times on average. The significant gain achieved even with the scalar implementation highlights the efficiency of our approach. 

Using our proposed ISA extension, MQX, we achieve an additional 2.1 times speedup over the AVX-512 implementation, resulting in a 66.9 times speedup over the best baseline. We observe that MQX performance begins to degrade at the NTT size of $2^{16}$. We hypothesize that this is the point at which the memory required for NTT exceeds the capacity of the L2 cache. As MQX accelerates computation, the kernel becomes memory-bound, and spilling beyond L2 leads to the observed slowdown. This hypothesis is supported by the calculation that, for an NTT size of $2^{15}$, each stage of NTT must hold about 1 MB of 128-bit integers; for a $2^{16}$-point NTT, this requirement doubles to 2 MB, exceeding the 1.28 MB per-core L2 cache on Intel Xeon. In contrast, the performance of AVX-512 remains relatively flat across all NTT sizes, as it continues to be compute-bound.

\bparagraph{AMD EPYC.} As shown in Figure~\ref{fig:ntt_amd}, both GMP and OpenFHE exhibit performance trends similar to those on Intel Xeon. On AMD EPYC, our scalar implementation achieves an average 11 times speedup over OpenFHE across all NTT sizes. Unlike on Intel Xeon, AVX2 outperforms the scalar implementation across all NTT sizes by an average of 1.2 times. AVX-512 delivers a further 1.7 times speedup over AVX2, resulting in a 23.2 times improvement over the state-of-the-art baseline, OpenFHE. The gains from MQX on AMD EPYC are even more evident than on Intel Xeon: With MQX, we achieve another 3.7 times speedup over AVX-512, yielding an overall 86.5 times improvement over OpenFHE.

\subsection{Sensitivity Analysis}
\label{sec:sa}

\begin{figure}[t]
\centering
    \includegraphics[width=0.95\columnwidth,trim={10mm 0mm 10mm 5mm},clip]{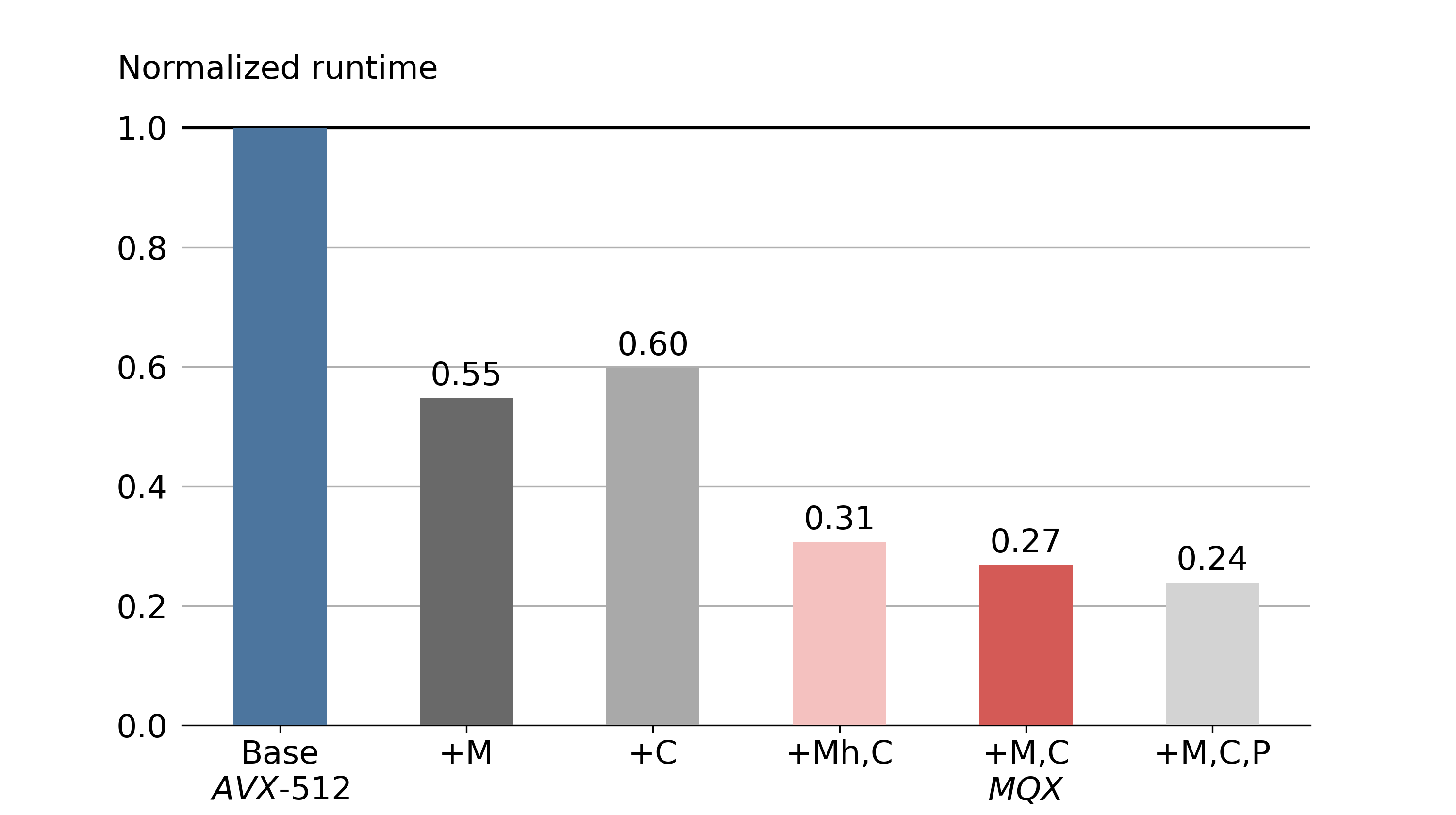}
    \caption{Sensitivity analysis of NTT runtime with respect to MQX instructions.}
    \label{fig:mqx_ablt}
\end{figure}

\bparagraph{MQX components.}
We quantify how each component of MQX contributes to the overall performance gain over the best-performing existing SIMD instruction set, AVX-512. In Figure~\ref{fig:mqx_ablt}, we report the average runtime per butterfly across all tested NTT sizes, normalized to the AVX-512 runtime (Base). All experiments were conducted on AMD EPYC.
In the figure, +M refers to augmenting AVX-512 with only widening multiplication, while +C adds only carry-flag support (including both addition with carry and subtraction with borrow). +M,C represents the full MQX extension, incorporating both features. The results show that widening multiplication contributes slightly more to performance improvements than carry-flag support when each is added individually.  

Next, we investigate whether the hardware requirements can be reduced by replacing full widening multiplication with a multiply-high operation, denoted as +Mh,C. Under our PISA model, we project the performance of widening multiplication using a single multiply-low instruction. In this analysis, we also model multiply-high with the same latency as multiply-low, decomposing the widening multiplication into two separate instructions.
Our results show that this substitution leads to only a minor performance degradation. This suggests that hardware vendors could implement a multiply-high instruction as a lower-cost alternative to full widening multiplication, while still retaining most of the performance benefits.

\bparagraph{Predicated execution.} 
When designing MQX, we also explored whether adding predicated execution support (denoted as +M,C,P) to the existing MQX instructions would further improve performance. To this end, we propose predicated addition with carry and predicated subtraction with borrow. Predicated addition with carry returns either the summation of two words and a carry-in as a single word (without setting the output carry flag), or simply returns the first input word. Predicated subtraction with borrow is defined similarly. 
Predication was also explored in LRBni, where each vector instruction can be predicated using a mask register. Similar concepts have been explored in the Intel Itanium architecture through the use of 1-bit predicate registers.
However, as shown in Figure~\ref{fig:mqx_ablt}, the performance improvement from predicated execution is modest---only a 1.1 times speedup over standard MQX. As a result, we chose not to include predicated executions in MQX in order to minimize additional hardware engineering effort.

\bparagraph{Choice of multiplication algorithms.} 
We conduct another sensitivity analysis to evaluate the performance impact of two multiplication algorithms, schoolbook and Karatsuba, across various NTT implementations. Our results show that schoolbook outperforms Karatsuba in almost all NTT variants (i.e., scalar, AVX2, AVX-512, and MQX) on both CPUs, with the only exception that their performance is nearly identical for the scalar implementation on AMD EPYC. When the schoolbook multiplication achieves better performance, it provides an average 1.1 times speedup over Karatsuba. These results suggest that, for NTT workloads on CPUs, replacing multiplication with more additions using Karatsuba does not lead to significant performance gains. This contrasts with prior findings on GPUs~\cite{zhang2025code}, where the Karatsuba algorithm achieved a 2.1 times speedup over the schoolbook multiplication for 128-bit inputs on NVIDIA GeForce RTX 4090. 

\begin{figure*}[t]
\centering
    \subfloat[Intel Xeon family.\label{fig:ntt_intel_sol}]
    {\centering\includegraphics[width=\columnwidth,trim={15mm 3mm 20mm 10mm},clip]{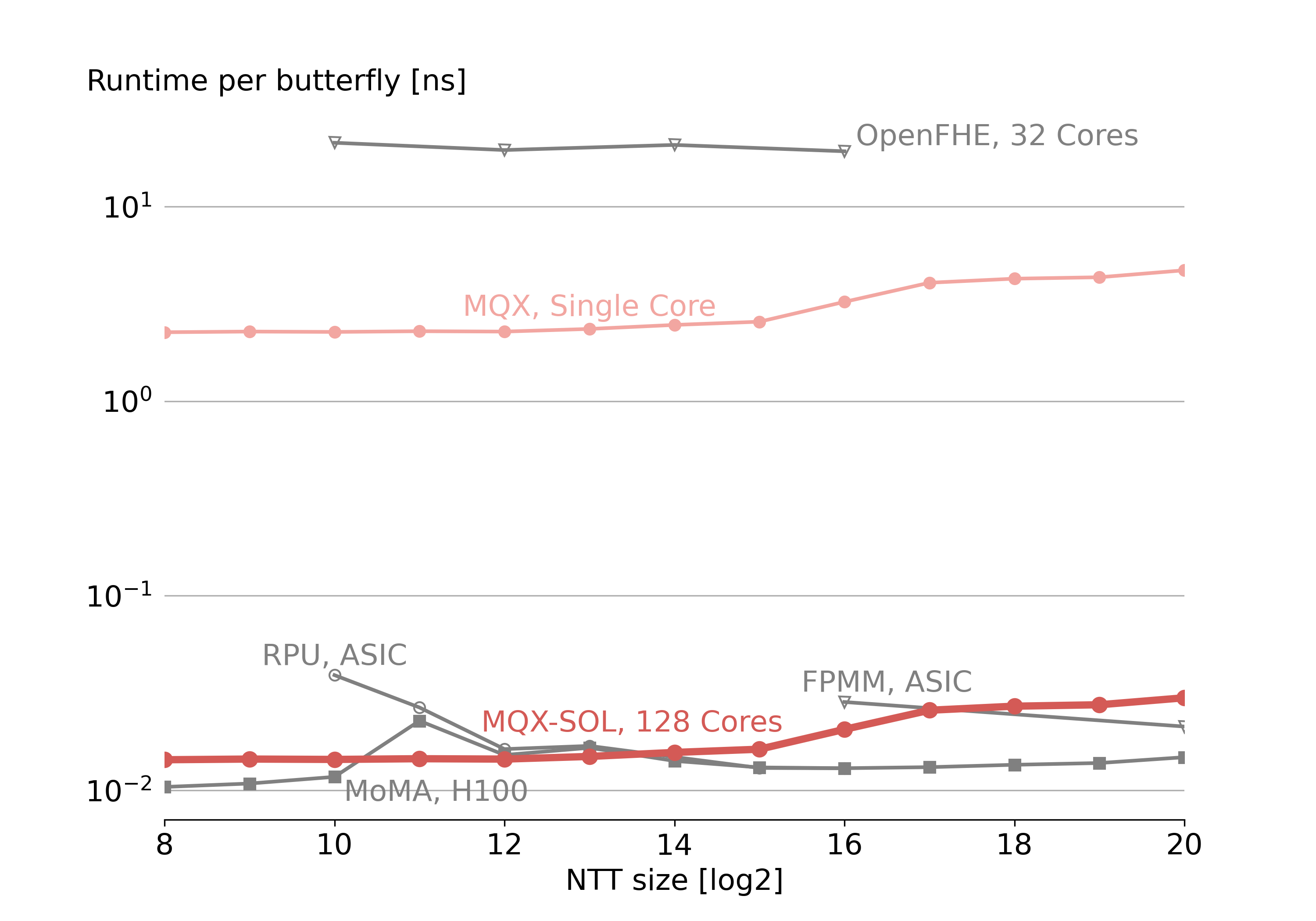}}
    \hspace{6mm}
    \subfloat[AMD EPYC family.\label{fig:ntt_amd_sol}]{\centering\includegraphics[width=\columnwidth,trim={15mm 3mm 20mm 10mm},clip]{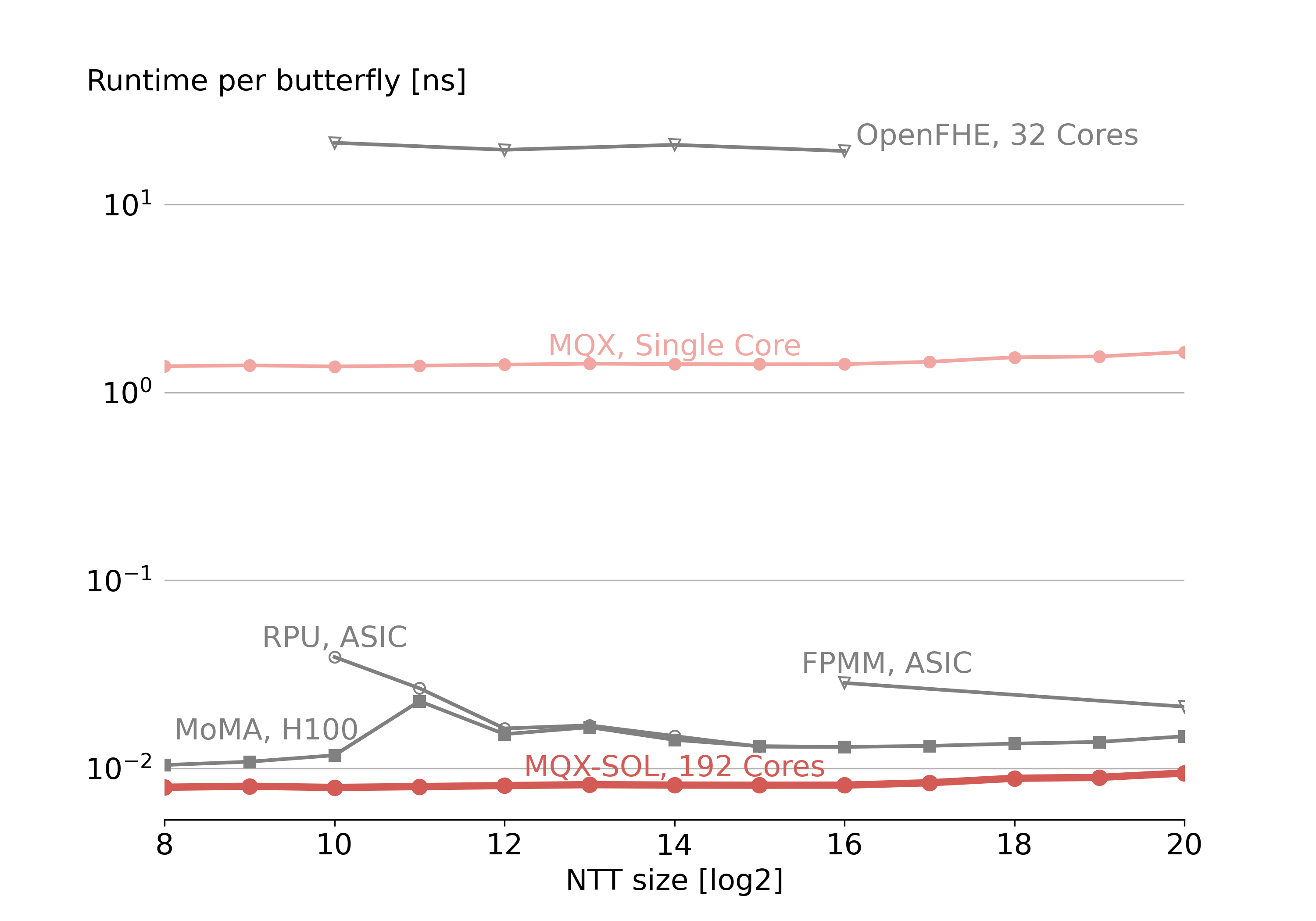}}
    \caption{Speed-of-light performance of NTT on multi-core CPUs.}
    \vspace{-3mm}
\end{figure*}

\section{Roofline Analysis}
\label{sec:sol}

In this section, we present a roofline-like~\cite{williams2009roofline} speed-of-light (SOL) performance model to estimate the upper bound achievable on top-tier server-class CPUs, assuming an optimal mapping of our current single-core implementation across the entire CPU. 
Our analysis compares projected CPU performance against two ASIC implementations, RPU~\cite{soni2023rpu} and Zhou et al.’s work~\cite{zhou2024fully}, one GPU implementation, MoMA~\cite{zhang2025code}, and the multi-core performance of OpenFHE, as reported in the RPU paper. For brevity, we henceforth refer to Zhou et al.’s work as FPMM.

For our SOL performance estimates, we select the highest-end server-grade Intel and AMD CPUs that support AVX-512. We normalize the performance by clock frequency and scale it by the number of cores. 
Formally, let $c_1$ and $c_2$ denote the number of cores on the CPU used for measurement and the target CPU, respectively. Let $f_\text{m}$ be the measured operating frequency and $f_\text{max}$ the all-core boost frequency of the target CPU. Given the measured runtime $t_\text{m}$, we define the SOL runtime $t_\text{sol}$ as 
\begin{equation}
\label{eq:sol}
  t_\text{sol} = t_\text{m} \cdot \frac{c_1}{c_2} \cdot \frac{f_\text{m}}{f_\text{max}}.
\end{equation}
Since all our measurements are taken on a single core, the equation above simplifies to $t_\text{sol} = t_\text{m} \cdot {f_\text{m}}/({c_2 \cdot f_\text{max}})$.

\bparagraph{Intel Xeon.}
The highest-end Intel CPU we choose within the Xeon family is Intel Xeon 6980P, which has 128 cores, 504 MB L3 cache, with an all-core boost frequency of 3.2 GHz. As shown in Figure~\ref{fig:ntt_intel_sol}, MQX-SOL is on average 1.3 times \textit{faster} than RPU across the NTT sizes supported by RPU, outperforming it at sizes from 1,024 to 8,192. We also include another ASIC baseline, FPMM~\cite{zhou2024fully}; averaging across the two NTT sizes that FPMM supports, MQX-SOL delivers approximately the same performance as FPMM. Compared to the GPU baseline, MoMA, MQX-SOL is 1.4 times slower on average.

\bparagraph{AMD EPYC.}
We choose AMD EPYC 9965S as the highest-end CPU for our speed-of-light analysis. AMD EPYC 9965S has 192 cores with an all-core boost frequency of 3.35 GHz, and 384 MB of L3 cache. Modeled on AMD EPYC 9965S, MQX-SOL achieves a 2.5 times \textit{speedup} over RPU on average across all RPU-supported NTT sizes and 2.9 times \textit{speedup} over FPMM. Compared to the GPU baseline MoMA, MQX-SOL achieves 1.7 times speedup on average. 

\bparagraph{Towards realizing SOL performance.}
We acknowledge that the MQX-SOL runtimes shown in the figures represent idealized projections and may not be fully achievable in practice, as near-linear scaling across all cores requires significant effort, including advanced thread management and low-level system optimization.
The SOL estimates intend to provide an upper bound when scaling our single-core implementation to an entire top-tier server-class CPU. Although optimistic, this model helps illustrate what could be possible under ideal conditions with careful system design and sufficient engineering effort. In practice, we can leverage the fact that real FHE workloads often batch NTTs and BLAS operations without data dependencies~\cite{al2022openfhe,zhang2025code}, enabling substantial parallelism. 

Nonetheless, the SOL analysis offers valuable insight into the performance potential of MQX. For example, our projected SOL performance on AMD EPYC 9965S outperforms RPU by 2.5 times. This suggests that, if we achieve a 77 times speedup by mapping our single-core implementation to batch NTTs on a 192-core machine, our performance would be on par with ASICs. Under a more conservative assumption of a 48 times speedup when mapping to a 192-core machine, our approach would be only about 1.6 times slower than RPU, demonstrating that near-ASIC performance is attainable on general-purpose CPUs.

It is also worth noting that ASIC and GPU measurements typically exclude data transfer time between the host CPU and the accelerator, whereas our CPU-based measurements include all data movement. This further strengthens our case: If the entire computation is hosted on a single CPU, the overhead of moving data between the host and the accelerator can be eliminated altogether.

\section{Discussion}
\label{sec:discussion}

\bparagraph{Automatic mapping to multi-core CPUs.}
In this work, we develop highly optimized implementations of cryptographic kernels for a single CPU core. 
One practical next step is to integrate the techniques proposed in this work with a compiler or code generation framework to automate the process of hardware-specific optimization and mapping to multiple CPU cores. This could be achieved by abstracting our hand-written implementation into an intermediate representation and integrating with existing compilation passes. For example, the high-performance code generator SPIRAL~\cite{franchetti2018spiral} has been used to generate optimized SIMD implementations of the fast Fourier transform, targeting instruction sets such as LRBni and AVX~\cite{mcfarlin2011automatic,franchetti2009discrete,franchetti2006fft}. Moreover, SPIRAL already supports code generation for NTT and BLAS operations~\cite{zhang2025code,zhang2023generating,zhang2023nttongpu}, making it a strong candidate for completing the last mile toward achieving SOL performance.

\bparagraph{Generalizing to larger bit-widths.} 
Another advantage of integrating this work with a compiler or code generation framework is the ability to generalize our optimized hand-written code to support higher bit-widths. The GPU baseline we compare against, MoMA, introduces a mathematical formalization of multi-word modular arithmetic and implements a rewrite system that can be combined with our approach. Leveraging MoMA, we can further extend our work to support higher input bit-widths, enabling its use in other cryptographic applications such as zero-knowledge proofs (ZKPs). This approach would offer a more performant alternative to general-purpose arbitrary-precision libraries like GMP for many cryptographic workloads on CPUs.

\bparagraph{Implementing MQX.} 
In this work, we estimate the potential performance of MQX and suggest possible implementation strategies (e.g., widening multiplication as a single instruction or as a sequence of multiply-low and multiply-high), leaving the implementation details open for future exploration. That said, we have shown that very similar instructions were proposed and implemented by Intel in the past, and our design follows Intel’s established methodology for introducing ISA sub-families (e.g., AVX-512 VNNI for neural networks and AVX-512 GFNI for Galois fields).
By identifying a minimal subset of the existing and legacy instructions and adapting them to a new, compelling domain---cryptography for privacy-preserving machine learning and artificial intelligence---we illustrate the performance potential that could be unlocked if a hardware vendor decides to implement MQX (or a simplified version of MQX, as demonstrated in Section~\ref{sec:sa}).

\section{Related Work}
\label{sec:related_work}

In this section, we first review prior work that utilizes specialized hardware and GPU for NTT acceleration, followed by a review of current solutions on CPUs.

\bparagraph{Specialized hardware and GPU for NTT acceleration.}
Due to the importance of NTT in FHE-based workloads, many specialized hardware accelerators have been developed for NTTs~\cite{samardzic2021f1,samardzic2022craterlake,soni2023rpu,wang2023sam,zhou2024fully,daftardar2024szkp,jayashankar2024cinnamon}.
However, few of these works target 128-bit NTTs. Among the designs that support 128-bit integer arithmetic, the performance gap between the CPU baseline and ASIC is substantial. For example, RPU~\cite{soni2023rpu} achieves a speedup of 545 to 1,485 times compared to the CPU baseline implemented using OpenFHE on a 32-core machine. Our work, even without MQX, reduces this gap to as low as 138 times on a single CPU core. This makes our work a stronger CPU baseline for future ASIC implementations.

There is also a significant body of work on GPU acceleration for NTTs~\cite{longa2016speeding,kim2020accelerating,durrani2021accelerating,ozerk2022efficient,shivdikar2022accelerating,wan2022novel,wang2023he,ozcan2023homomorphic,livesay2023accelerating,wang2023nttfusion,zhang2025code}. These works primarily focus on bit-widths that are natively supported by GPUs, due to the lack of performant multi-precision libraries for GPUs. A recent work, MoMA~\cite{zhang2025code}, proposes and implements a rewrite system that can take arbitrary input bit-widths and decompose them into machine-supported arithmetic, achieving near-ASIC performance on commodity GPUs. One potential future direction is to integrate our CPU optimizations into MoMA and incorporate them into compilers or code generators to target higher input bit-widths. Similar to our MQX proposal, Shivdikar et al.~\cite{shivdikar2023gme} propose coarse-grained modular instructions for accelerating FHE on AMD compute DNA (CDNA) GPUs, embedding full modular operations (e.g., modular multiplication and reduction) into single high-level vector instructions. In contrast, MQX introduces fine-grained SIMD extensions for x86 CPUs, with each instruction closely resembling existing or legacy SIMD patterns to support multi-word modular arithmetic.

\bparagraph{Current solutions on CPU.}
Current solutions on CPUs are primarily FHE libraries such as OpenFHE~\cite{al2022openfhe}, SEAL~\cite{chen2017seal}, HEXL~\cite{boemer2021intel}, HElib~\cite{helib}, and TFHE~\cite{chillotti2020tfhe}. Most of these libraries are highly generalizable, supporting multiple FHE schemes. Notably, Intel's HEXL~\cite{boemer2021intel} utilizes AVX-512 to accelerate FHE-based workloads but only for 64-bit arithmetic. 
A closely related work by van der Hoeven and Lecerf~\cite{van2024implementing} also uses SIMD instructions such as AVX2 and AVX-512 to accelerate NTT on CPUs, but their approach relies on modular floating-point arithmetic and a specialized prime to accelerate computations, while our method works with general primes.

Similar to most GPU-based approaches, the majority of CPU-based solutions support only 32-bit or 64-bit arithmetic and rely on RNS to break down large integers into these natively supported bit-widths. OpenFHE provides a built-in mathematical library for 128-bit integer arithmetic; however, its implementation is 32 times slower than our AVX-512 implementation on Intel Xeon.
Another popular approach to implement NTT with large input bit-widths is to use arbitrary-precision libraries like GMP~\cite{granlund1996gnu}. Libraries such as a library for doing number theory (NTL)~\cite{shoup2001ntl} and fast library for number theory (FLINT)~\cite{hart2013flint} also support arbitrary-precision arithmetic, but rely on GMP as the backend. In Section~\ref{sec:ntt_results}, we show that our AVX-512-based NTT outperforms the GMP baseline by 53 times on Intel Xeon.

\section{Conclusion}

To answer the question of whether near-specialized hardware performance can be achieved on CPUs for cryptographic kernels, our work begins by using scalar and SIMD instructions to accelerate NTTs and BLAS operations, resulting in 38 times and 62 times speedup over state-of-the-art CPU baselines, respectively. 
We further enhance performance by proposing MQX, an ISA extension that addresses the performance bottlenecks of AVX-512 for large-integer arithmetic. MQX introduces only three new SIMD instructions, each grounded in similar instructions that were previously explored and implemented by Intel.
Using a roofline analysis, we demonstrate that by optimally scaling the single-core performance of MQX across all cores of a server-grade CPU, we can achieve near-ASIC performance for all tested NTT sizes. 
This finding suggests that, with continued software optimizations and minimal changes to the hardware ISA, CPUs have the potential to match the performance of GPUs and ASICs, making them a viable alternative for cryptographic workloads.

\begin{acks}
The authors thank Eric Tang for many helpful discussions throughout the course of this work.
This material is based upon work supported by the National Science Foundation under Grant No. 1127353, the U.S. Department of Energy, Office of Science, Office of Advanced Scientific Computing Research under Award Numbers DE-FOA-0002460 and DE-SC0025645, and PRISM, a center in JUMP 2.0, a Semiconductor Research Corporation (SRC) program sponsored by the Defense Advanced Research Projects Agency (DARPA). 
This work used FASTER and Launch at Texas A\&M University through allocation CIS230287 from the Advanced Cyberinfrastructure Coordination Ecosystem: Services \& Support (ACCESS) program, which is supported by National Science Foundation grants \#2138259, \#2138286, \#2138307, \#2137603, and \#2138296.
Any opinions, findings, and conclusions or recommendations expressed in this material are those of the authors and do not necessarily reflect the views of the National Science Foundation, the U.S. Department of Energy, and DARPA.
Franz Franchetti was partially supported as the Kavčić-Moura Professor of Electrical and \mbox{Computer Engineering}. 
\end{acks}

\appendix
\section{Artifact Appendix}

\subsection{Abstract}

Our artifact includes the source code for our highly optimized implementation of BLAS operations and NTTs, targeting CPUs with AVX2 and AVX-512 support. Compilation on Intel CPUs uses the Intel oneAPI DPC++/C++ Compiler (ICX), while compilation on AMD CPUs uses the AMD Optimizing C/C++ and Fortran Compiler (AOCC), which employs Clang as its frontend.

\subsection{Artifact Checklist (Meta-Information)}

{\small
\begin{itemize}
  \item {\bf Compilation: } ICX $\geq$ 2024.2.0, AOCC $\geq$ 4.1.0, and Clang $\geq$ 16.0.3. 
  \item {\bf Dataset: } The required dataset is included with this artifact.
  \item {\bf Hardware: } Intel and AMD CPUs as detailed in Table~\ref{tab:cpus}.
  \item {\bf Runtime state: } Users should ensure that no other processes are competing for system resources during benchmarking. 
  \item {\bf Execution: } We provide a detailed \verb|README| file within the artifact, and users only need to run the provided Bash scripts to perform benchmarking.
  \item {\bf Metrics: } Execution time.
  \item {\bf Output: } Performance measurements will be displayed in
the terminal window.
  \item {\bf How much disk space is required (approximately)?: } Around 6 MB.
  \item {\bf How much time is needed to complete experiments (approximately)?: } Less than 1 hour. 
\end{itemize}
}

\subsection{Description}

Here we provide a short description of how the artifact is delivered and its dependencies. 

\bparagraph{How delivered.} The artifact is provided as a repository on GitHub\footnote{\url{https://github.com/naifeng/benchntt}}. The artifact requires approximately 6 MB of disk space.

\bparagraph{Hardware dependencies.} To reproduce the exact results, Intel Xeon 8352Y and AMD EPYC 9654 are required. Otherwise, the hardware should at least support AVX2 and AVX-512 instructions.

\bparagraph{Software dependencies.} For Intel CPUs, the artifact requires ICX version 2024.2.0 or later for compilation. For AMD CPUs, it requires AOCC version 4.1.0 or later, with AMD Clang version 16.0.3 or later.

\subsection{Installation}

Instructions for installing dependencies are provided in the \verb|README| file.

\subsection{Experiment Workflow}

A detailed \verb|README| file is provided in the root directory of the artifact. For example, to reproduce Figure~\ref{fig:blas_amd}, run the following command on an AMD CPU:
\begin{lstlisting}[style=shellblock]
 $ bash ./benchmark/blas_aocc.sh
\end{lstlisting}
\vspace{3mm}
To reproduce Figure~\ref{fig:ntt_intel}, run the following command on an Intel CPU:
\begin{lstlisting}[style=shellblock]
 $ bash ./benchmark/ntt_icx.sh
\end{lstlisting}

\subsection{Evaluation and Expected Results}

After the experiments are complete, the terminal will display the results as follows: for NTT, the runtime per butterfly for each NTT size; for each BLAS operation, the runtime per element. All runtimes are reported in nanoseconds.

\subsection{Experiment Customization}

Users can customize the parameters in Equation~\ref{eq:sol} to match their specific CPUs for more accurate roofline analysis. Detailed instructions are provided in the \verb|README| file.


\bibliographystyle{ACM-Reference-Format}
\bibliography{ref}

\end{document}